\documentclass[showpacs,showkeys,preprintnumbers,prd,amsmath,amssymb,floats,floatfix]{revtex4}
\usepackage{amsmath,amssymb,bm,natbib}
\usepackage[paperwidth=210mm, paperheight=297mm,centering,hmargin=3cm,vmargin=3.5cm]{geometry}
\usepackage{epstopdf}
\usepackage{hyperref}
\usepackage{color}
\usepackage{booktabs}
\usepackage{graphicx}
\usepackage{multirow}
\usepackage{slashed}
\usepackage{comment}
\usepackage{float}
\makeatletter

\newcommand{\Rmnum}[1]{\expandafter\@slowromancap\romannumeral #1@}

\makeatother
\begin{document}
\def\bib{\bibitem}
\def\be{\begin{equation}}
\def\ee{\end{equation}}
\def\beq{\begin{equation}}
\def\eeq{\end{equation}}
\def\beqar{\begin{eqnarray}}
\def\eeqar{\end{eqnarray}}
\def\barr{\begin{array}}
\def\earr{\end{array}}
\def\dis{\displaystyle}
\def\lsim{\:\raisebox{-0.5ex}{$\stackrel{\textstyle<}{\sim}$}\:}
\def\gsim{\:\raisebox{-0.5ex}{$\stackrel{\textstyle>}{\sim}$}\:}
\def\tilh{\tilde{h}}
\def\and{\qquad {\rm and } \qquad}
\def\vev{\small \em {\it v.e.v. }}
\def\p{\partial}
\def\ga{\gamma^\mu}
\def\slp{p \hspace{-1ex}/}
\def\sleps{ \epsilon \hspace{-1ex}/}
\def\slk{k \hspace{-1ex}/}
\def\slq{q \hspace{-1ex}/\:}
\def\prd#1{Phys. Rev. {\bf D#1}}
\def\etal{ {\it et al.} }
\def\ie{ {\it i.e.} }
\def\eg{ {\it e.g.} }

\title{Study of Higgs-gauge boson anomalous couplings through $e^-e^+ \rightarrow W^-W^+h$ at the ILC}
\author{\bf Satendra Kumar} \email{satendra@iitg.ernet.in} 

\affiliation{
Department of Physics,
Indian Institute of Technology Guwahati,
Guwahati 781 039, India}

\author{\bf P. Poulose} \email{poulose@iitg.ernet.in}

\affiliation{
Department of Physics,
Indian Institute of Technology Guwahati,
Guwahati 781 039, India}

\author{\bf Shibananda Sahoo} \email{shibananda@iitg.ernet.in} 

\affiliation{
Department of Physics,
Indian Institute of Technology Guwahati,
Guwahati 781 039, India}

\begin{abstract} 
Higgs couplings with gauge bosons are probed through $e^-e^+ \rightarrow W^-W^+h$ in an effective Lagrangian framework. An ILC of 500 GeV center-of-mass energy with possible beam polarization is considered for this purpose. The reach of the ILC with integrated luminosity of 300 fb$^{-1}$ in the determination of both the CP-conserving and CP-violating parameters is obtained. Sensitivity of the probe of each of these couplings on the presence of other couplings is investigated. The most influential coupling parameters are $\bar c_W=-\bar c_B$. Other parameters of significant effect are $\bar c_{HW}$ and $\bar c_{HB}$ among the CP-conserving ones, and $\tilde c_{HW}$  among the CP-violating ones. CP-violating parameters, $\tilde c_\gamma$ and $\tilde c_{HB}$ are found to have very little influence on the process considered. A detailed study of the angular distributions presents a way to disentangle the effect of some of these couplings.
\end{abstract} 
\maketitle
\newpage
\section{Introduction}
The discovery of the new resonance of mass around 126 GeV by the ATLAS and CMS collaborations at the LHC \cite{cms, atlas, Moriond1, Moriond2, Moriond3, Moriond4, Moriond5,Aad:2013wqa,Chatrchyan:2012ufa,Chatrchyan:2013lba,Aad:2012tfa,Khachatryan:2014ira,Aad:2014eha} provides a gateway to the 
investigations of the dynamics of elementary particles. The discovery has unambiguously established the role of the Higgs mechanism in electroweak symmetry breaking (EWSB). All of the properties of the new particle measured so far are consistent with that of the standard Higgs boson. Thus, one may be tempted to conclude that for all practical purposes, the newly found particle is like that of the Standard Model (SM) Higgs boson, and new physics effects are decoupled as far as the Higgs sector is concerned. At the same time, it is well known that there are difficulties associated with the Higgs sector of the SM that need to be addressed. The main difficulty is the hierarchy problem associated with the quadratically diverging quantum corrections to the mass of the Higgs boson when computed in the SM. There is no remedy to this difficulty within the SM, and for a Higgs boson of mass 126 GeV, the new physics effects should show up within the TeV range to cure this malady. Assuming that the new physics effects are expected to appear only indirectly in the Higgs sector, it is natural to consider these effects through effective couplings of the Higgs bosons, with itself as well as with the gauge bosons and heavy fermions. Precise measurement of these couplings is very essential to establish the true nature of the EWSB mechanism. While the LHC is capable of probing some of these couplings \cite{Gabrielli:2013era}, especially the Higgs couplings with the gauge bosons and top quark, one may need to rely on a cleaner machine like the International Linear Collider (ILC) \cite{ILC1,ILC2,polarizationreview,Boos:2014xza} for the required precision. Another aspect that is very important to investigate is the CP properties of the couplings of the Higgs boson. Although the measurements so far indicate a CP-even Higgs boson, it is not ruled out that the Higgs sector does not involve any CP violation. One may remember that, one of the compelling reasons to look beyond the SM is the large CP-violation necessary to understand the baryon asymmetry of the Universe. There have been many studies on the CP properties of the Higgs boson in the past \cite{HiggsCP}. More recently there have been studies on the CP properties of the Higgs interaction with the top quark \cite{Ananthanarayan:2014eea}, investigating the influence of a CP-mixed Higgs boson on the Yukawa couplings. Within an effective Lagrangian, the effect of new physics could be studied in the various couplings through the quantum corrections they acquire. Such an effective Lagrangian basically encodes the new physics effects in higher-dimensional operators with anomalous couplings. 

The study of the Higgs sector through an effective Lagrangian goes back to Refs.\cite{Weinberg:1978kz,Weinberg:1980wa,Georgi:1994qn,Buchmuller:1985jz,Hagiwara:1993ck,Hagiwara:1993qt,Alam:1997nk,
genuined6,Giudice:2007fh,Contino2010a,Contino2010b,Grober2011,Grzadkowski:2010es}.
 More recently,  the Lagrangian including a complete set of dimension-six operators was studied by 
Refs. \cite{Baak:2012kk, Einhorn:2013kja,Contino:2013kra,Willenbrock:2014bja}. For some of the recent references discussing the constraints on the anomalous couplings within different approaches, please see Refs.
\cite{Bonnet:2011yx,Corbett:2012dm,Chang:2013cia,Elias-Miro:2013mua,Banerjee:2013apa,Boos:2013mqa,
Masso:2012eq,Han:2004az, Corbett:2012ja,Dumont:2013wma, Pomarol:2013zra,Ellis:2014dva, Belusca-Maito:2014dpa, Gupta:2014rxa}.
Reference \cite{Ellis:2014dva} studied the $h+V$, where $V= Z, W,$ associated production at the LHC and Tevatron to discuss the bounds obtainable from the global fit to the presently available data, whereas Ref.~\cite{Belusca-Maito:2014dpa} has discussed the constraint on the parameters coming from the LHC results as well as other precision data from LEP, SLC, and Tevatron. 
Experimental studies on the Higgs couplings at the LHC are presented in, for example, Refs. \cite{ 1307.1427,Teyssier:2014hta}.
The measurement of trilinear Higgs couplings is best done through the process $e^+ e^-\rightarrow Zhh$  \cite{De Rujula:1991se,1,2,3,4,5,6,7,8,9,10}.  At the same time, this process also depends on the Higgs-gauge boson couplings, $ZZh$ and $ZZhh$, which will affect the determination of the $hhh$ coupling. Another process that could probe the $hhh$ couplings is $e^+e^- \rightarrow \nu_e \bar\nu_e hh$ following the $WW$ fusion \cite{4,5,6,7}, which is also affected by the $WWh$ and $WWhh$ couplings. In a recent study \cite{Kumar:2014zra}, we investigated the effect of the $VVh$ coupling, where $V=Z, ~W$, in the extraction of  the $hhh$ coupling, and found that a precise knowledge of the $WWh$ and $ZZh$ couplings is necessary to derive information regarding the trilinear couplings. 

The process $e^+ e^-\rightarrow W^-W^+h$ is well suited to study the Higgs to gauge boson couplings \cite{De Rujula:1991se,1,2,3,4,5,6,7,8,9,10}. At the same time, this process also depends on the trilinear gauge boson couplings like $WW\gamma$, which can contaminate the effects of Higgs to gauge boson couplings. In this paper we will focus our attention on this process in some detail within the framework of the effective Lagrangian. One goal of this study is to investigate CP violation in the Higgs sector through Higgs to gauge boson couplings and to understand the significance of other couplings in their measurement. 
 
The paper is presented in the following way. In Sec.~\ref{sec:setup}, the effective Lagrangian will be presented, with the currently available constraint on the parameters. In Sec.~\ref{sec:discussions}, the process under consideration will be presented, with details. In Sec.~\ref{sec:summary}, the results will be summarized. 

\section{General Setup}\label{sec:setup}
 References \cite{Giudice:2007fh,Contino2010a,Contino2010b,Grober2011,Contino:2013kra,Ellis:2014dva, HEL} present the most general effective Lagrangian with dimension-six operators involving the Higgs bosons. Part of this Lagrangian relevant to the process $e^+ e^-\rightarrow W^-W^+h$ considered in this paper is given by
 
\begin{eqnarray}
{\cal L}_{\rm Higgs}^{\rm CPC} &=&    \frac{\bar c_{H}}{2 v^2} \partial^\mu\big(\Phi^\dag \Phi\big) \partial_\mu \big( \Phi^\dagger \Phi \big) +\frac{\bar c_{T}}{2 v^2} \big(\Phi^\dag \overleftrightarrow D^\mu \Phi\big) \big( \Phi^\dagger \overleftrightarrow D_\mu \Phi \big) -  \frac{\bar c_6}{v^2}\lambda~\big(\Phi^\dag\Phi\big)^3+\frac{\bar c_{\gamma}}{m_W^2} g'^2~\Phi^\dag \Phi B_{\mu\nu} B^{\mu\nu}
  \nonumber\\
  &&+ \frac{i g ~\bar c_{HW}}{m_W^2}~\big(D^\mu \Phi^\dag \sigma_{k} D^\nu \Phi\big) W_{\mu \nu}^k 
  + \frac{i g' ~\bar c_{HB}}{m_W^2}~ \big(D^\mu \Phi^\dag D^\nu \Phi\big) B_{\mu \nu} \nonumber\ \\
  && + \frac{ i g ~\bar c_{W}}{2m_W^2} ~\big( \Phi^\dag \sigma_{k} \overleftrightarrow{D}^\mu \Phi \big)  D^\nu  W_{\mu \nu}^k 
  +\frac{i g' ~\bar c_{B}}{2 m_W^2} ~\big(\Phi^\dag \overleftrightarrow{D}^\mu \Phi \big) \partial^\nu  B_{\mu \nu},
   \nonumber\ \\[5mm]
{\cal{L}}^{\rm CPV} &=& \frac{ig ~\tilde c_{HW}}{m_W^2} D^\mu \Phi^\dag T_{2k} D^\nu \Phi \tilde W^k_{\mu\nu} + \frac{ig'~ \tilde c_{HB}}{m^2_W} D^\mu \Phi^\dag D^\nu \Phi \tilde B_{\mu\nu}  \nonumber\\
&&+  \frac{{g'}^2~ \tilde c_\gamma}{m^2_W} \Phi^\dag \Phi B_{\mu\nu} \tilde B_{\mu\nu} + \frac{g^3 ~\tilde c_{3W}}{m^2_W} \epsilon_{ijk} W^i_{\mu\nu} W_\rho^{\nu j} \tilde W^{\rho \mu k}
\label{eq:Leff}
\end{eqnarray}
where the dual field strength tensors are defined as 
\( \tilde B_{\mu\nu} = \frac{1}{2}\epsilon_{\mu\nu\rho\sigma} B^{\rho\sigma},~~~\tilde W^k_{\mu\nu} = \frac{1}{2}\epsilon_{\mu\nu\rho\sigma} W^{\rho\sigma k}
\nonumber
\)
and
\(
  \Phi^\dag {\overleftrightarrow D}_\mu \Phi = 
    \Phi^\dag (D_\mu \Phi) - (D_\mu\Phi^\dag) \Phi \ ,
\) with $D_\mu$ being the appropriate covariant derivative operator and $\Phi$ the usual Higgs doublet in the SM. Also, $W_{\mu\nu}^k$ and $B_{\mu\nu}$ are the field tensors corresponding to the $SU(2)_L$ and $U(1)_Y$ of the SM gauge groups, respectively, with gauge couplings  $g$ and $g'$, in that order. $\sigma_k$ are the Pauli matrices, and $\lambda$ is the usual (SM) quadratic coupling constant of the Higgs field.
The above Lagrangian leads to the following CP-conserving (${\cal L}_{hV} ^{\rm CPC}$) and CP-violating ($ {\cal L}_{hV} ^{\rm CPV}$) parts in the unitary gauge and mass basis~\cite{HEL}:

\begin{eqnarray}
 \label{eq:LagPhys}
  {\cal L}_{hV} ^{\rm CPC} &=& -\frac{1}{4} g_{hzz}^{(1)} Z_{\mu\nu} Z^{\mu\nu} h-
  g_{hzz}^{(2)} Z_{\nu}\partial_{\mu} Z^{\mu\nu} h 
  + \frac{1}{2} g_{hzz}^{(3)} Z_{\mu} Z^{\mu} h 
  -\frac{1}{2} g_{h\gamma z}^{(1)} Z_{\mu\nu} F^{\mu\nu} h \nonumber \\
  &&-g_{h\gamma z}^{(2)}Z_{\nu} \partial_{\mu} F^{\mu\nu} h 
 - \frac{1}{8} g_{hhzz}^{(1)}Z_{\mu\nu} Z^{\mu\nu} h^2 - \frac{1}{2} g_{hhzz}^{(2)}Z_{\nu} \partial_{\mu} Z^{\mu\nu} h^2 + \frac{1}{4} g_{hhzz}^{(3)}Z_{\mu} Z^{\mu} h^2 \nonumber\\
%
 &&- \frac{1}{2} g_{hww}^{(1)} W^{\mu\nu} W_{\mu\nu}^{\dagger} h -\left [g_{hww}^{(2)} W^{\nu} \partial^{\mu} W_{\mu\nu}^{\dagger} h + H.c.\right]
 + g~m_{W} W_{\mu}^{\dagger} W^{\mu} h \nonumber \\
 && - \frac{1}{4} g_{hhww}^{(1)} W^{\mu\nu} W_{\mu\nu}^{\dagger} h^2 - \frac{1}{2} \left[g_{hhww}^{(2)} W^{\nu} \partial^{\mu} W_{\mu\nu}^{\dagger} h^2 + H.c.\right] + \frac{1}{4} g^2 W_{\mu}^{\dagger} W^{\mu} h^2\\[5mm]
  {\cal L}_{3V} &=& \Big[ig^{(1)}_{\gamma ww} W^\dag _{\mu\nu}A^{\mu}W^{\nu} + H.c. \Big] + i g^{(2)}_{\gamma ww} F_{\mu\nu}W^{\mu}W^{\nu \dag} \nonumber\\
&&+ \Big[ig^{(1)}_{zww}W^\dag_{\mu\nu}Z^\mu W^{\nu} + H.c. \Big] + ig^{(2)}_{zww}Z_{\mu\nu}W^{\mu}W^{\nu\dag}
\\[5mm]
  {\cal L}_{hV} ^{\rm CPV}&=& -\frac{1}{4} \tilde g_{h\gamma\gamma} F_{\mu\nu}\tilde F^{\mu\nu} h -\frac{1}{4} \tilde g_{hzz} Z_{\mu\nu}\tilde Z^{\mu\nu} h - \frac{1}{2} \tilde g_{h\gamma z} Z_{\mu\nu} \tilde F^{\mu\nu} h - \frac{1}{2} \tilde g_{hww} W^{\mu\nu} \tilde W^\dag_{\mu\nu} h \nonumber\\
&& - \frac{1}{8} \tilde g_{hh\gamma\gamma} F_{\mu\nu}\tilde F^{\mu\nu}h^2 - \frac{1}{8} \tilde g_{hhzz} Z_{\mu\nu}\tilde Z^{\mu\nu}h^2 -  \frac{1}{8} \tilde g_{hh\gamma z} Z_{\mu\nu}\tilde F^{\mu\nu}h^2 \nonumber\\
 &&- \frac{1}{4} \tilde g_{hhww} W^{\mu\nu}\tilde W^{\dag}_{\mu\nu}h^2 + i\tilde g^{(1)}_{hzww} \tilde Z^{\mu\nu} W_{\mu}W^\dag_{\nu}h - \left[i\tilde g^{(2)}_{hzww} \tilde W^{\mu\nu} Z_{\mu}W^\dag_{\nu}h + H.c.\right]~~~~~~~
\label{eq:LagPhysCPV}
\end{eqnarray}

\begin{table}[H]
\begin{tabular}{ll}
\hline 
&\\
CP-conserving couplings\\[2mm]
$g_{hzz}^{(1)}=\frac{2g}{c_W^2 m_W} \left[ \bar{c}_{HB} s_W^2 - 4\bar{c}_{\gamma} s_W^4 + c_W^2 \bar{c}_{HW}  \right]$&\\
$g_{hzz}^{(2)} =\frac{g}{c_W^2 m_W} \left[ (\bar{c}_{HW}+ \bar{c}_W) c_W^2 + (\bar{c}_B + \bar{c}_{HB}) s_W^2  \right]$&$g_{hzz}^{(3)}  = \frac{g m_Z}{c_W} \left[ 1-2 \bar{c}_T  \right]$\\
$g_{hz\gamma}^{(1)}  = \frac{g s_W}{c_W m_W} \left[ \bar{c}_{HW} -\bar{c}_{HB} + 8\bar{c}_{\gamma} s_W^2 \right]$&
$g_{hz\gamma}^{(2)}  = \frac{g s_W}{c_W m_W} \left[ \bar{c}_{HW} -\bar{c}_{HB} -\bar{c}_B + \bar{c}_W \right]$\\
$g_{hww}^{(1)}  = \frac{2g}{m_W} \bar{c}_{HW},$&$g_{hww}^{(2)}  = \frac{g}{2m_W} \left[\bar{c}_{W} + \bar{c}_{HW}\right]$\\
$g_{\gamma ww}^{(1)}  = e\big[1-2\bar c_W \big],$&$g_{\gamma ww}^{(2)}  = e\big[1-2\bar c_W - \bar c_{HB} - \bar c_{HW}\big]$\\
$g_{zww}^{(1)}  = \frac{g}{c_W}\big[c^2_W - \bar c_{HW} + (2 s^2_W - 3)\bar c_W]$\\
$g_{zww}^{(2)}  = \frac{g}{c_W}\big[c_W^2 (1-\bar c_{HW} )+ s^2_W \bar c_{HB} + (2 s^2_W -3)\bar c_{W}\big]$\\[5mm]
\hline 
\hline
&\\
CP-violating couplings\\[2mm]
 $\tilde g_{h\gamma\gamma}  = -\frac{8g \tilde {c}_{\gamma} s^2_W}{m_W}$,&$
\tilde g_{hz\gamma}  = \frac{gs_W}{c_W m_W} \left[\tilde {c}_{HW} - \tilde{c}_{HB} + 8 \tilde c_{\gamma}s^2_W\right]$\\
 $ \tilde g_{hzz}  = \frac{2g}{c^2_W m_W} \left[\tilde {c}_{HW}s^2_W -4\tilde{c}_{\gamma} s^4_W+c^2_W \tilde c_{HW}\right]$&$\tilde g_{hww}  = \frac{2g}{m_W} \tilde {c}_{HW}$\\
 $ \tilde g^{(1)}_{hzww}  = \frac{g^2}{c_Wm_W}\left[\tilde {c}_{HW}(2-s^2_W) + \tilde c_{HB}s^2_W\right]$&$
 \tilde g^{(2)}_{hzww}  = \frac{2g^2}{m_W}c_W \tilde {c}_{HW}$\\[5mm]
 \hline
\end{tabular}
\caption{Physical couplings in Eqs. (\ref{eq:LagPhys})-(\ref{eq:LagPhysCPV}) are given in terms of the effective couplings in Eq. (\ref{eq:Leff}), where $c_W = \cos\theta_W$ and $s_W = \sin\theta_W$, with $\theta_W$ being the weak mixing angle .}
\label{table:couplings}
\end{table}

The physical couplings relevant to the process $e^+e^-\rightarrow WWh$, and associated with the Lagrangian in Eqs.~(\ref{eq:LagPhys}) - (\ref{eq:LagPhysCPV}) expressed in terms of the effective couplings presented in Eq.~(\ref{eq:Leff}) are listed in Table~\ref{table:couplings}.
In total, there are nine parameters which are relevant to the process considered, viz,  $\bar{c}_T, \bar{c}_{\gamma}, \bar{c}_B, \bar{c}_W, \bar{c}_{HB}, \bar{c}_{HW}, \tilde{c}_{HW}, \tilde{c}_{HB}, \tilde{c}_{\gamma} $. Out of these, six parameters are related to CP-conserving couplings, and the other three are connected with CP-violating couplings. 
These anomalous coefficients $\bar{c}_T, \bar{c}_{HW}, \bar{c}_{HB}, \bar{c}_{\gamma}$  are expected to be of the order
\begin{equation}
 \bar{c}_T \sim \mathcal{O}\left( \frac{g_{\sc NP}^2 v^2}{M^2}\right) ~~{\rm and}~~ 
 \bar{c}_{HW}, \bar{c}_{HB}, \bar{c}_{\gamma} \sim \mathcal{O}\left( \frac{{g_{\sc NP}^2} M_{W}^2}{16 \pi^2 M^2}\right),
\end{equation}
where $g_{\sc NP}$ denotes the generic coupling of the new physics, and $M$ is the new physics scale. This indicates that
these couplings can be significantly large for strongly coupled physics. In contrast the coefficients of the operators
such as $\bar{c}_W$ and $\bar{c}_B$ are given by
\begin{equation}
 \bar{c}_B, \bar{c}_W \sim \mathcal{O}\left( \frac{m_W^2}{M^2}\right)
\end{equation}
and therefore, expected to be relatively suppressed or enhanced according to the ratio ${g/g_{\sc NP}}$. Coming to the experimental bounds,  electroweak precision data put the following constraints~\cite{Baak:2012kk},
\begin{equation}
 \bar{c}_T(m_Z) \in [-1.5, 2.2] \times 10^{-3} ~{\rm and}~ (\bar{c}_W(m_Z) + \bar{c}_B(m_Z))\in [-1.4, 1.9]\times 10^{-3}
 \label{eq:cTcWcBconst}
\end{equation}
This means we can safely ignore the effect of $\bar c_T$ in our analysis. On the other hand, $\bar{c}_W$ and $\bar{c}_B$ are not independently constrained, leaving the possibility of having  large values with a cancellation between them as per the above constraint. $\bar c_W$ itself along with $\bar c_{HW}$ and $\bar c_{HB}$ are constrained from LHC observations on the associated production of the Higgs along with $W$ in Ref. ~\cite{Ellis:2014dva}.
Considering the Higgs-associated production along with $W$, ATLAS and CMS along with D0 put a limit of  \( \bar{c}_W \in \big[-0.03, 0.01\big] \), when all other parameters were set to zero. A global fit using various information from ATLAS and CMS including signal-strength information constrains the region in the $\bar c_W-\bar c_{HW}$ plane, leading to a slightly more relaxed limit on $\bar c_W$ and a limit of about \( \bar{c}_{HW} \in \big[-0.1, 0.06\big] \).  The limit on $\bar{c}_{HB}$ estimated using a global fit in Ref.~\cite{Ellis:2014dva} is about \(\bar{c}_{HB} \in [-0.05, 0.05]\) with a one-parameter fit. The CP-violating couplings are largely unconstrained so far. 

The purpose of this study is to understand how to exploit a precision machine like the ILC to investigate a suitable process so as to derive information regarding these couplings. In the next section, we shall explain the process of interest in the present case and discuss the details to understand the influence of one or more of the couplings mentioned above.

\section{Analyses of the process considered} \label{sec:discussions}
\begin{figure}[H] \centering
\begin{tabular}{c c}
\hspace{1mm}
\includegraphics[angle=0, width=90mm]{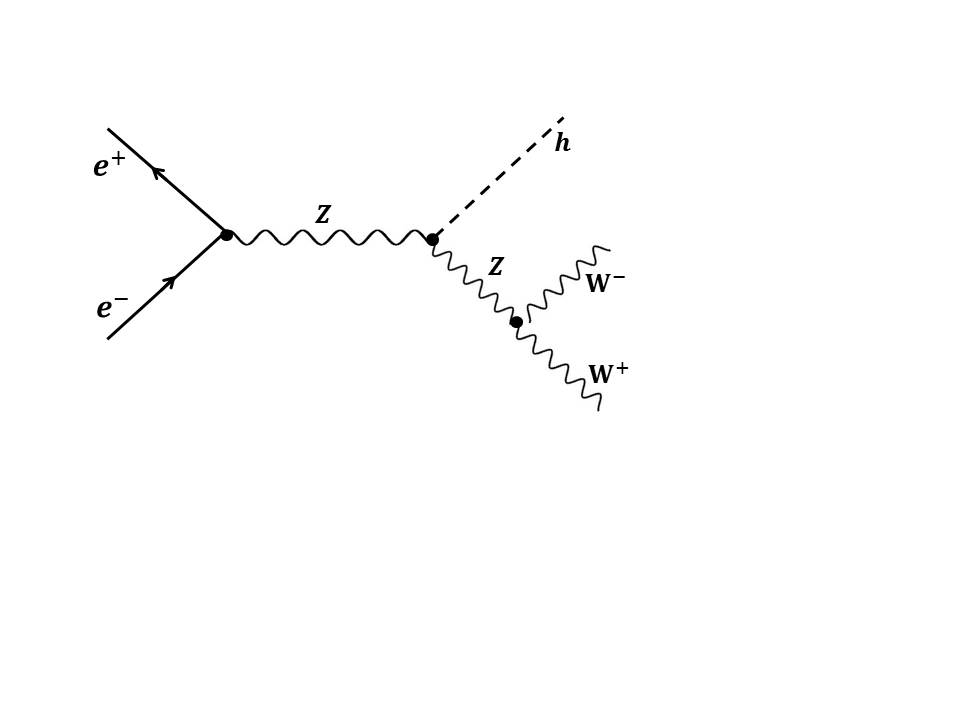} &
\vspace{-3cm}\hspace{-25mm}
\includegraphics[angle=0, width=90mm]{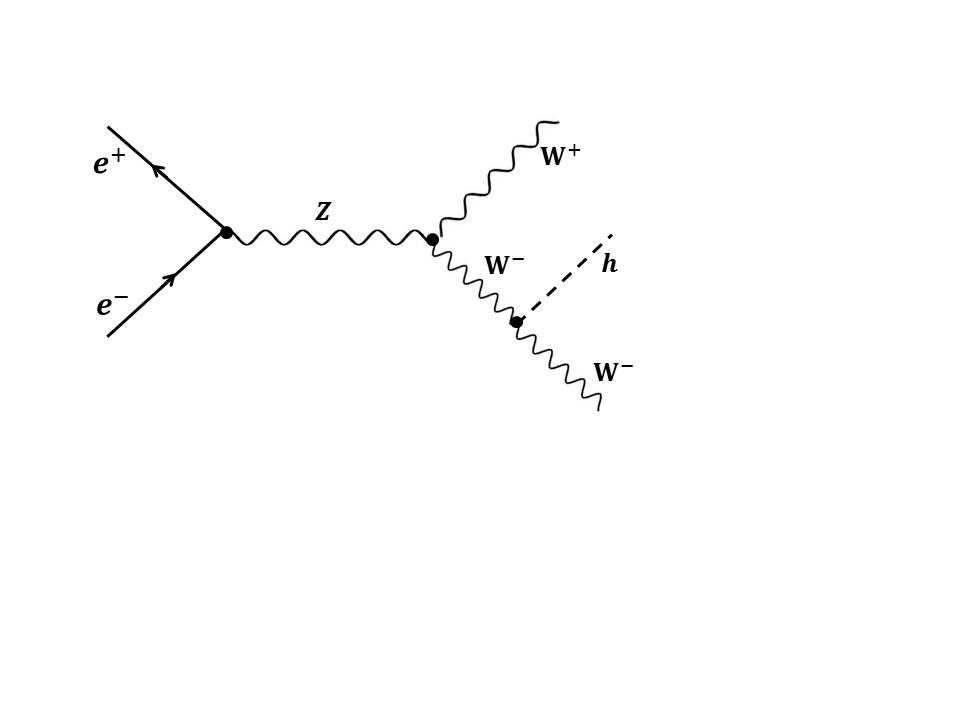} \\
\hspace{1mm}
\includegraphics[angle=0, width=90mm]{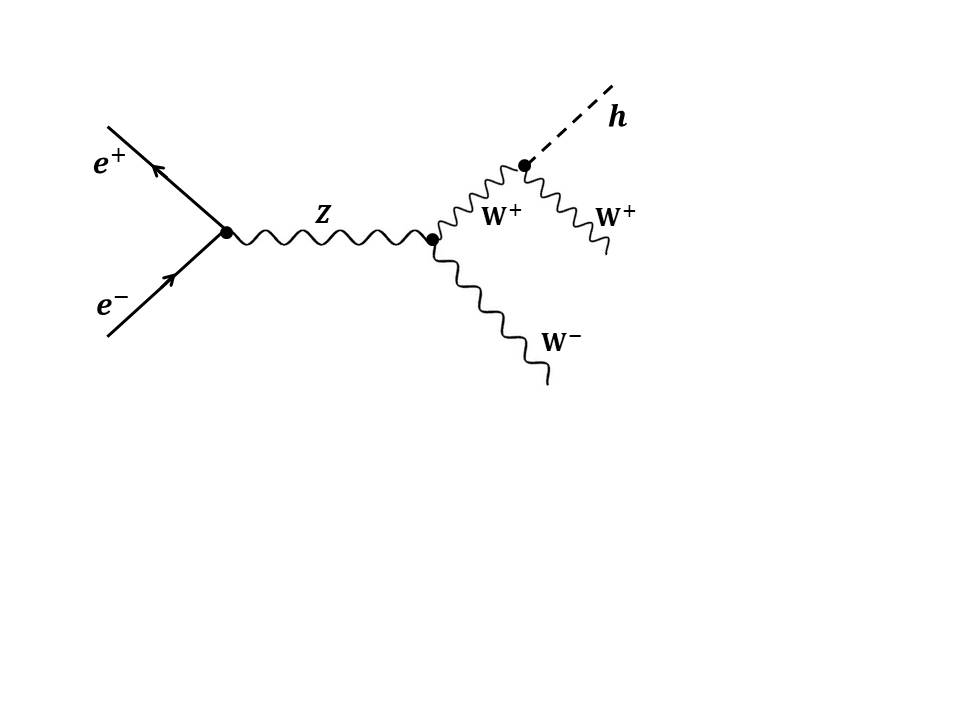} &
\vspace{-3cm}\hspace{-25mm}
\includegraphics[angle=0, width=90mm]{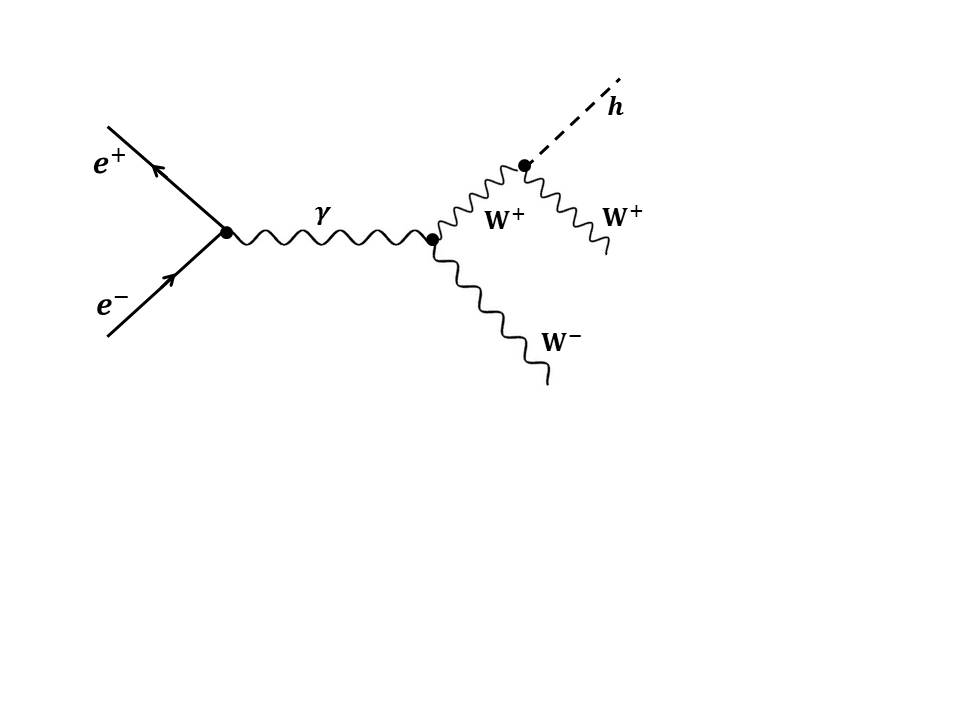} \\
\hspace{1mm}
\includegraphics[angle=0, width=90mm]{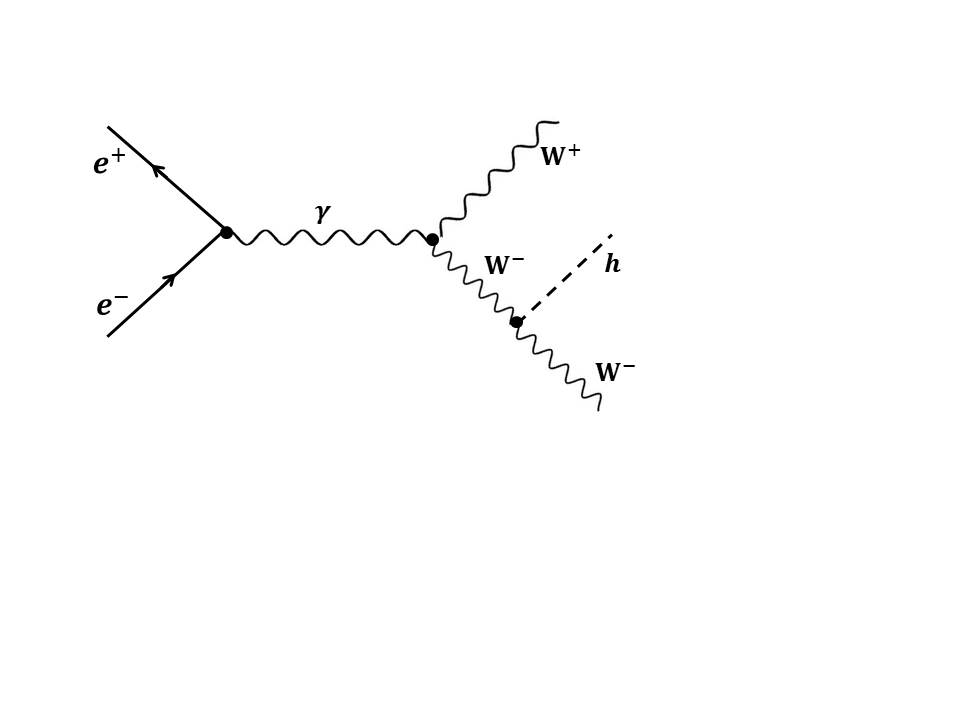} &
\vspace{-3cm}\hspace{-40mm}
\includegraphics[angle=0, width=90mm]{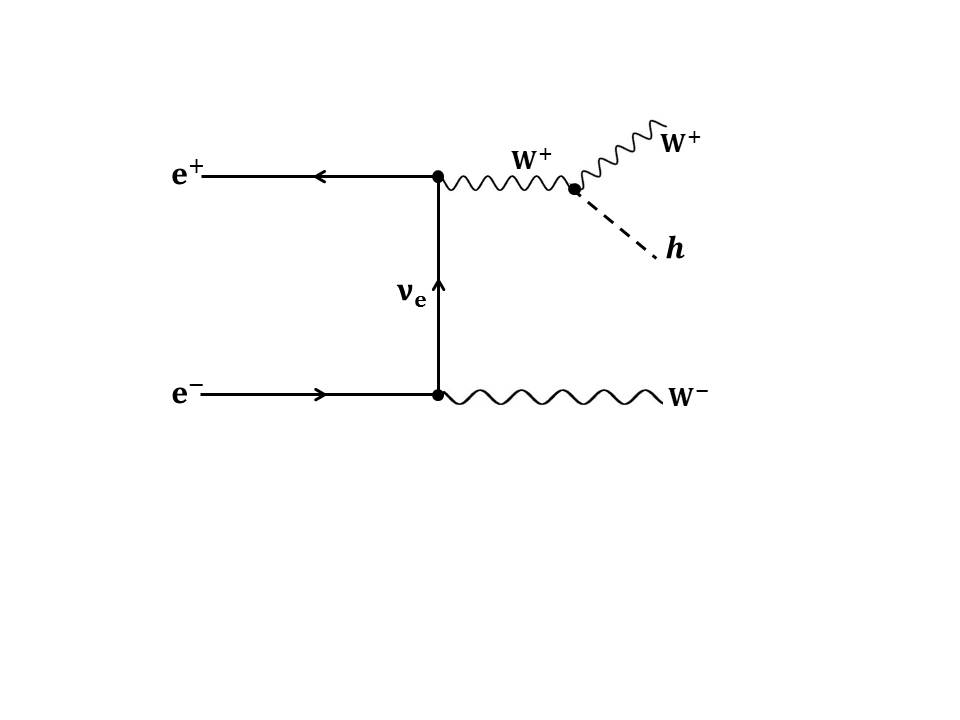} \\
\vspace{-2cm}\hspace{-5mm}
\includegraphics[angle=0, width=90mm]{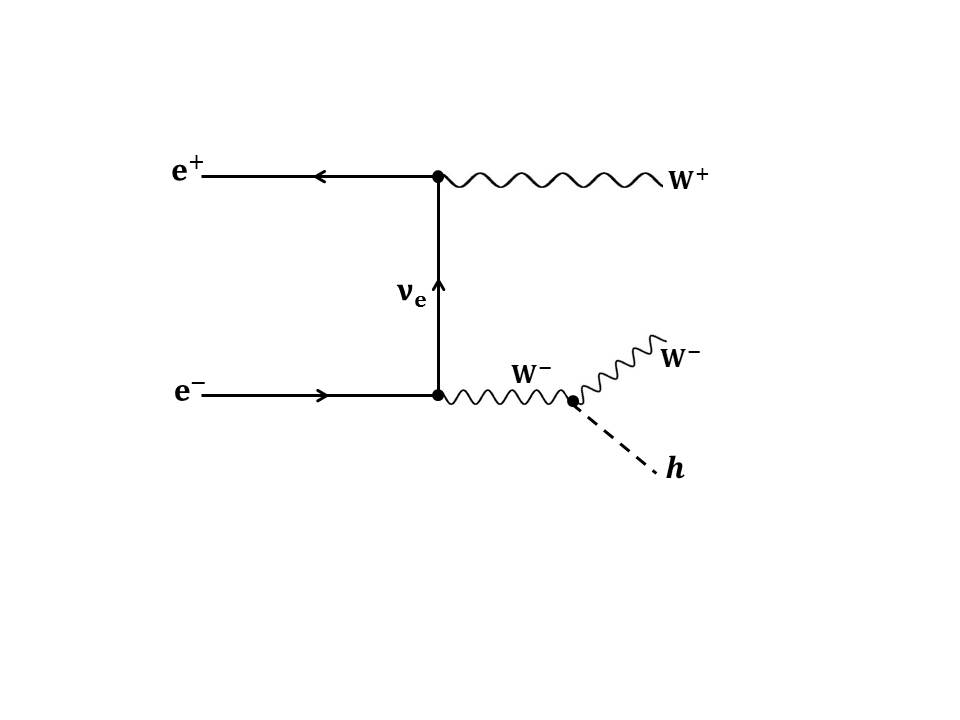} 
\hspace{-70mm}
\end{tabular}
\caption{Feynman diagrams contributing to the process $e^- e^+ \rightarrow W^- W^+ h$ in the SM.}
\label{fig:feynmandiag1}
\end{figure}

The Feynman diagrams corresponding to the process $e^- e^+ \rightarrow W^-W^+h$  in the SM are given in Fig.\ref{fig:feynmandiag1}. This process is basically influenced by Higgs to charged gauge boson as well as neutral gauge boson couplings like $WWh,~ZZh, ~ WW\gamma$, and $WWZ$, apart from the fermionic couplings, which are taken to be the standard couplings in our study. 

The effective Lagrangian, Eq. (\ref{eq:Leff}), apart from allowing the existing Higgs and gauge boson couplings to be nonstandard, introduces new couplings which are absent in the SM. In a specific model such effects appear at higher orders with a new particle present in the loops. When the masses of such particles are taken to be large, the effect of such quantum correction can be considered in terms of changed couplings. Such effective couplings arising in the present analysis are presented in Table~\ref{table:couplings}. 
Our numerical analyses are carried out using {\sc madgraph} \cite{madgraph, Alwall:2014hca}, with the effective Lagrangian implemented through {\sc feynrules} \cite{feynrules, HEL}.

\begin{figure}[H] \centering
\begin{tabular}{c c}
\hspace{-10mm}
\includegraphics[angle=0,width=90mm]{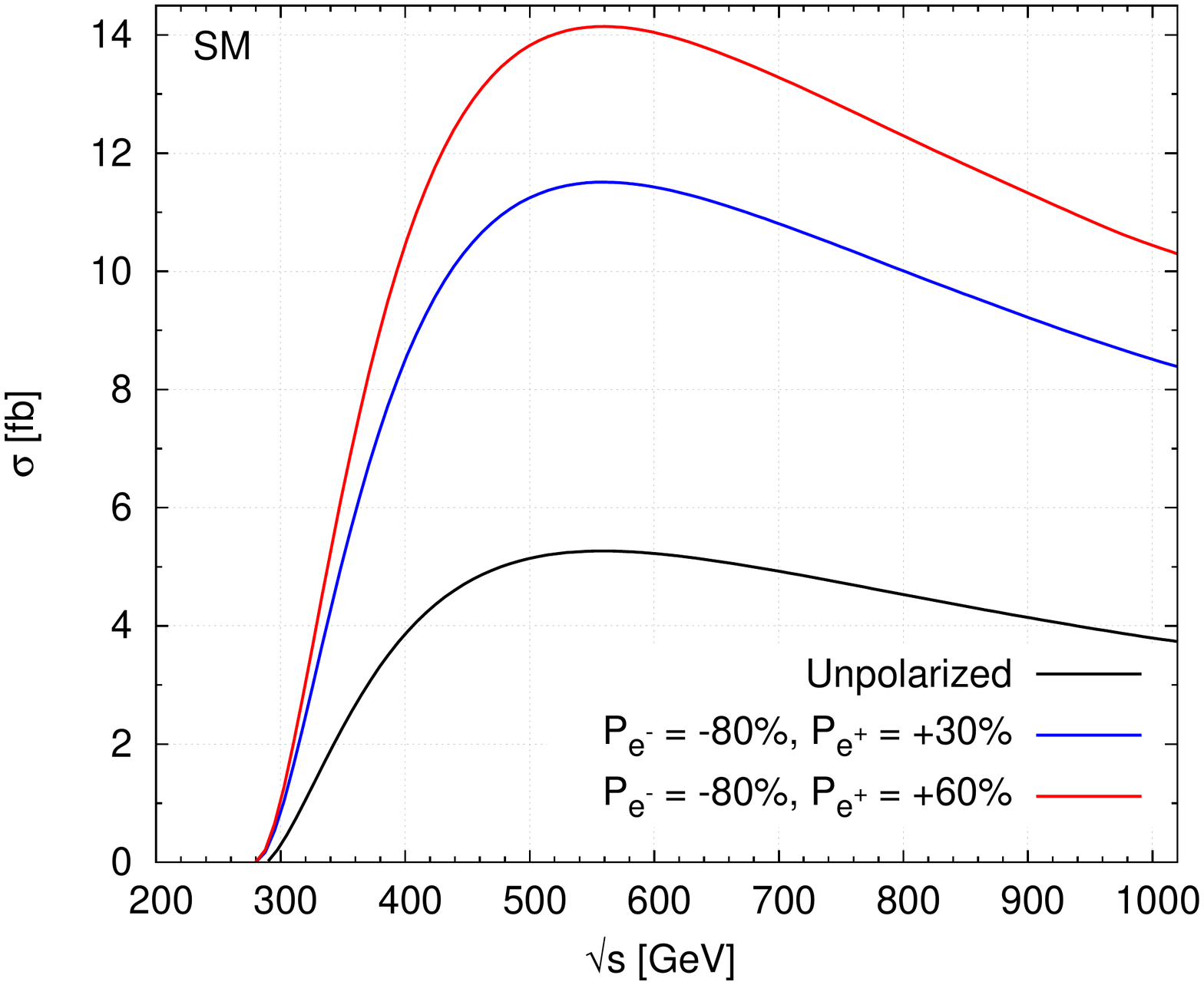} &
\hspace{-16mm}
\includegraphics[angle=0,width=90mm]{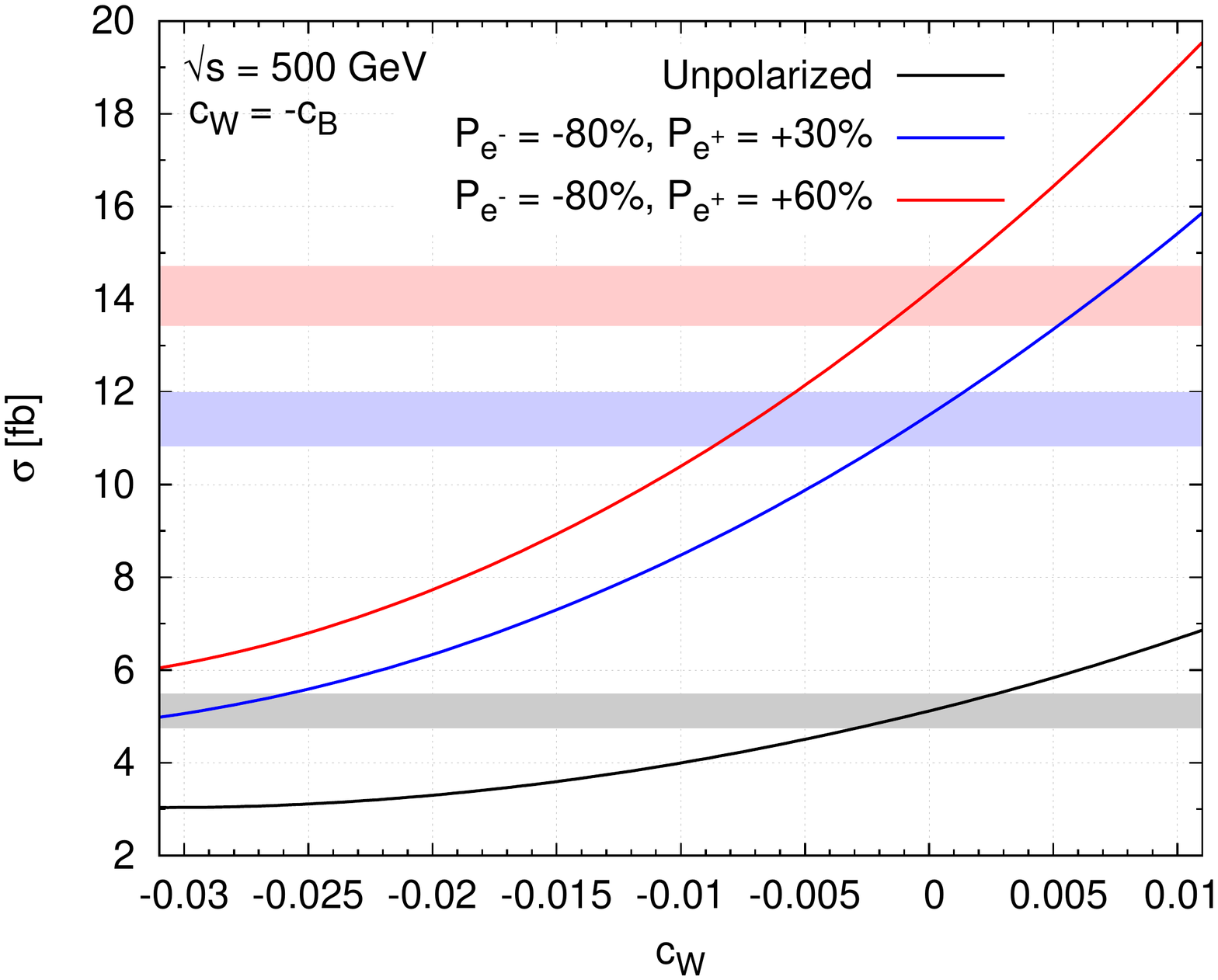} 
\end{tabular}
\vspace{-10mm}
\caption{{\bf Left}: The total cross section against $\sqrt{s}$ in the SM. {\bf Right}: The total cross section against anomalous coupling parameter ($\bar c_W$) at $\sqrt{s}=500$ GeV, where the gray, blue, and red bands correspond to $3\sigma$ deviations from the SM with unpolarized and polarized beams, respectively. Note that here, as well as in all figures henceforth, we have removed the ``bar" from the symbols denoting the CP-conserving parameters  for convenience.}
\label{fig:cs1}
\end{figure}

As the first observable, we consider the cross section. Figure \ref{fig:cs1} (left) presents the total cross section against the center-of-mass energy for the $WWh$ production. The cross section peaks around the center-of-mass energy of $500$ GeV, and, therefore, our further detailed analysis will be done for a collider of this energy. As expected, the polarization hugely improves the situation. The case of a typical polarization combination expected at the ILC, 80\% left-polarized electron beam and 30\% right-polarized positron beam, is considered \cite{polarizationreview}, along with the case of an 80\% left-polarized electron beam and a 60\% right-polarized positron beams, which are expected in the upgraded version of the ILC. In Fig.\ref{fig:cs1} (right) the cross section against an anomalous couplings parameter ($\bar c_W$) at fixed center-of-mass energy of $500$ GeV is considered along with the role of the polarized beams. In order to be consistent with the experimental constraint [Eq.~(\ref{eq:cTcWcBconst}], we choose $\bar c_B=-\bar c_W$ throughout our analysis. Notice that the cross section is enhanced rapidly, even for the very small values of $\bar c_W$, showing the high sensitivity of the cross section on this parameter. Assuming that no other couplings affect the process, the single parameter reach corresponding to the $3\sigma$ limit with $300~fb^{-1}$ integrated luminosity is obtained as $-0.003 \le (\bar c_W=-\bar c_B) \le +0.003$ in the case of unpolarized beam, which is improved to $-0.002 \le (\bar c_W=-\bar c_B) \le +0.002$ with an 80\% left-polarized electron beam and a 30\% right-polarized positron beam. While the case with an 80\% left-polarized electron beam and a 60\% right-polarized positron beam does not change this limit significantly, the cross section is increased from about 11 fb to about 14 fb, enhancing the statistics.  In our further analysis, we consider the baseline expectation of an 80\% left-polarized electron beam and a 30\% right-polarized positron beam.
 
\begin{figure}[H] \centering
\begin{tabular}{c c}
\hspace{-10mm}
\includegraphics[angle=0,width=90mm]{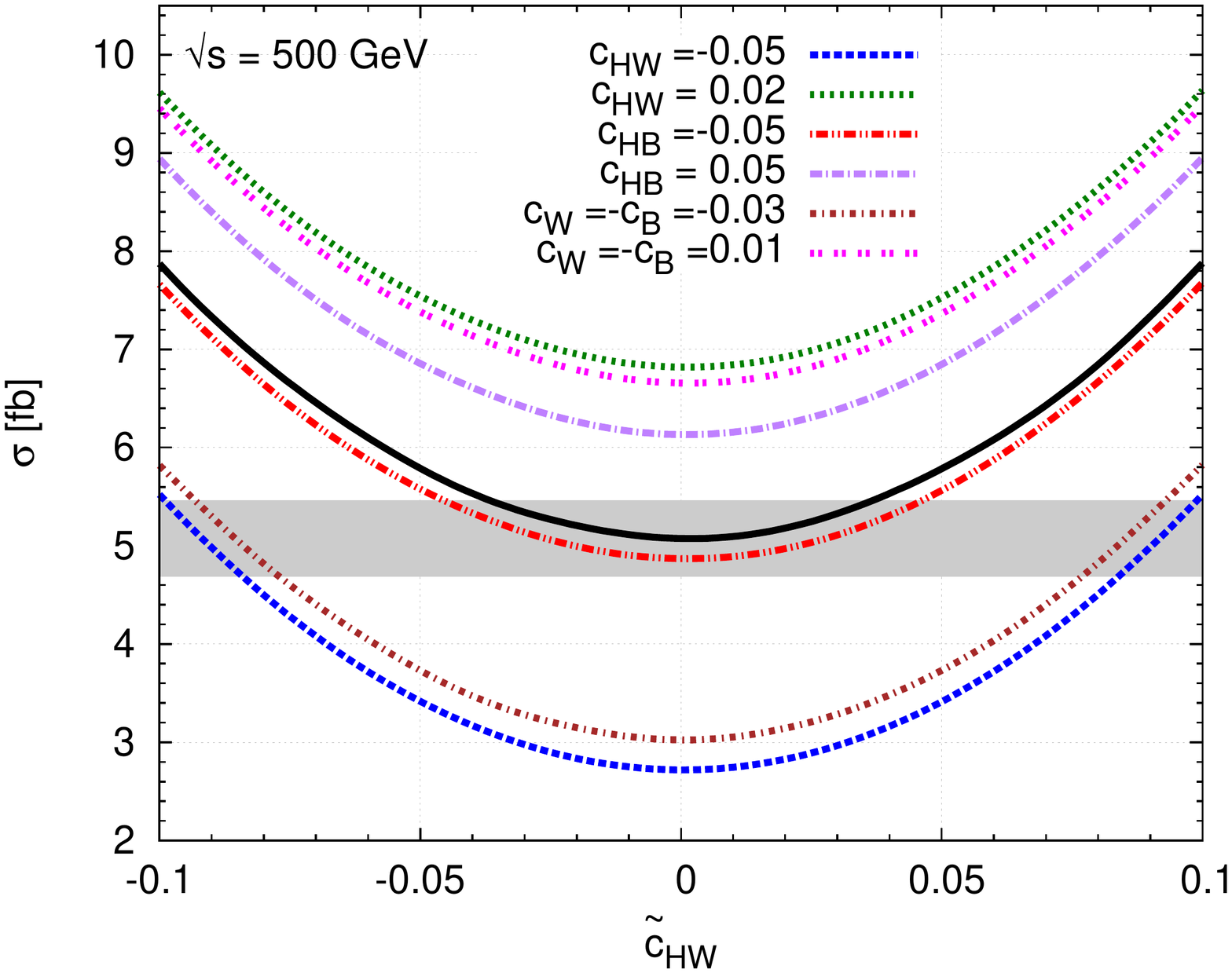} & 
\hspace{-18mm}
\includegraphics[angle=0,width=90mm]{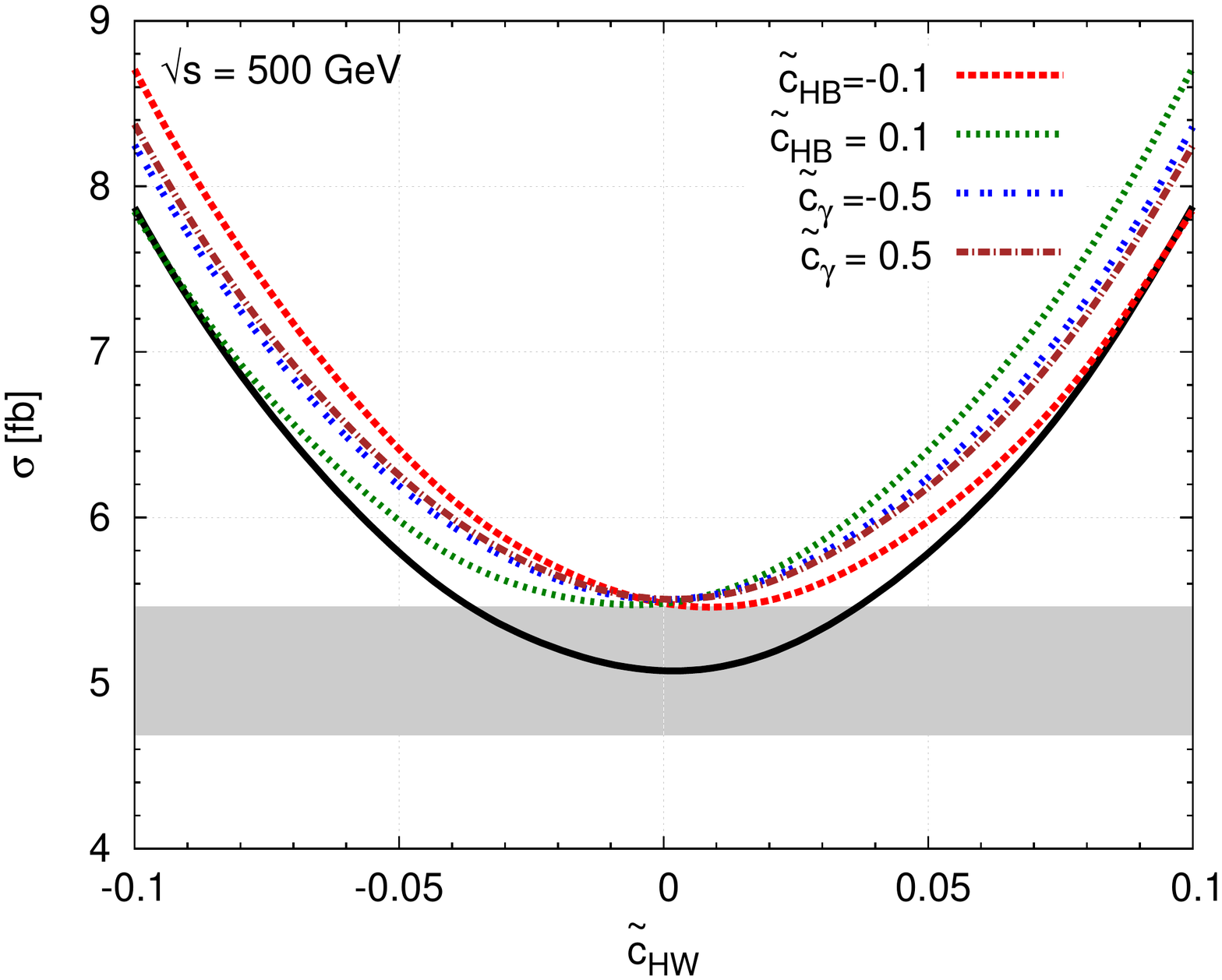}\\
\end{tabular}
\vspace{-10mm}
\caption{Cross section against $\tilde c_{HW}$ in the presence of selected CP-conserving (left) and CP-violating (right) couplings.
The black solid line corresponds to the case when only $\tilde c_{HW}$ is present. The center-of-mass energy is assumed to be 
$\sqrt{s}=500$ GeV. In each case, all other parameters are set to zero. The gray band indicates the $3\sigma$ limit 
of the SM cross section, with an integrated luminosity of $300$~fb$^{-1}$.}
\label{fig:cstcHW}
\end{figure}

\begin{figure}[H] \centering
\begin{tabular}{c c}
\hspace{-10mm}
\includegraphics[angle=0,width=90mm]{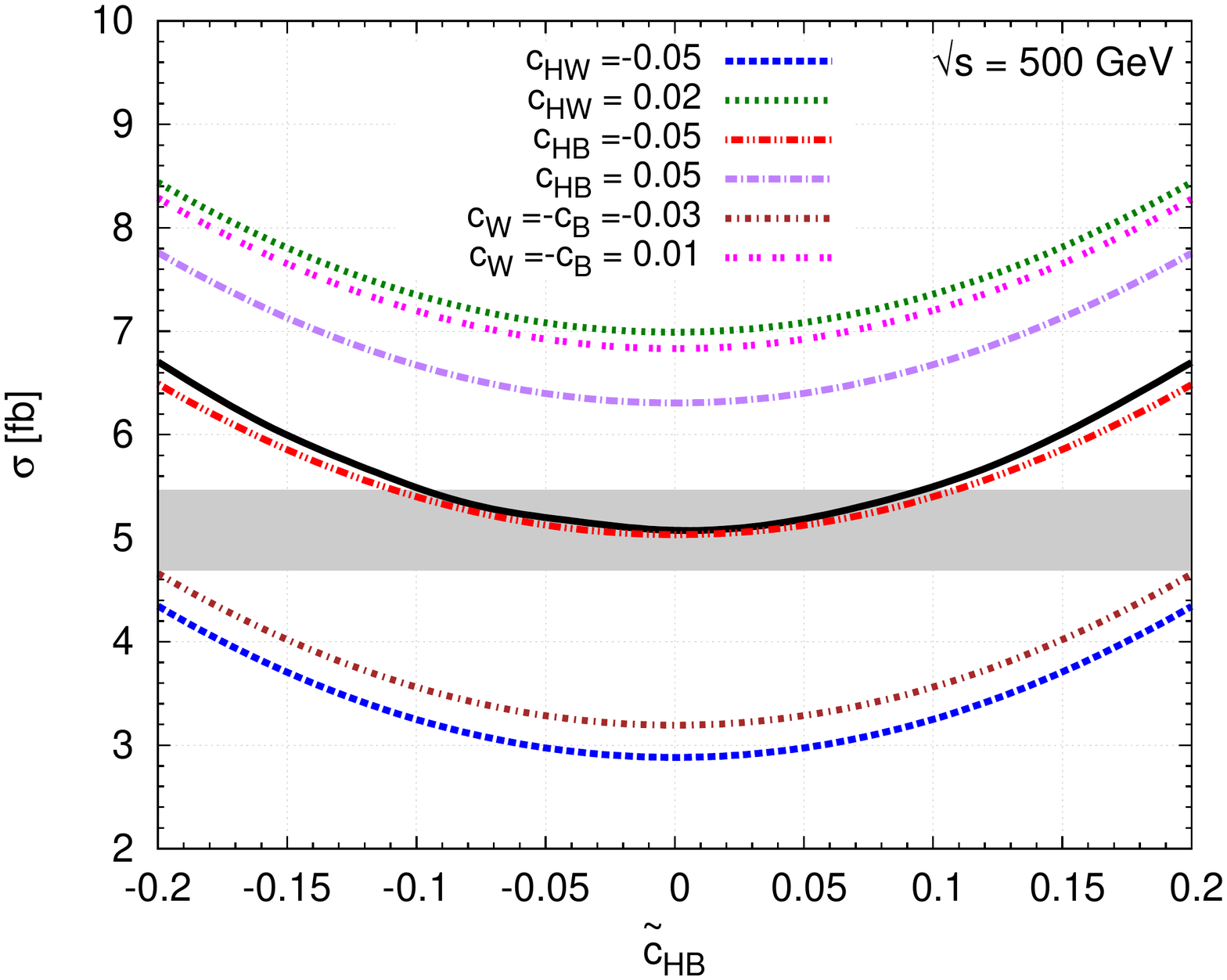} & 
\hspace{-16mm}
\includegraphics[angle=0,width=90mm]{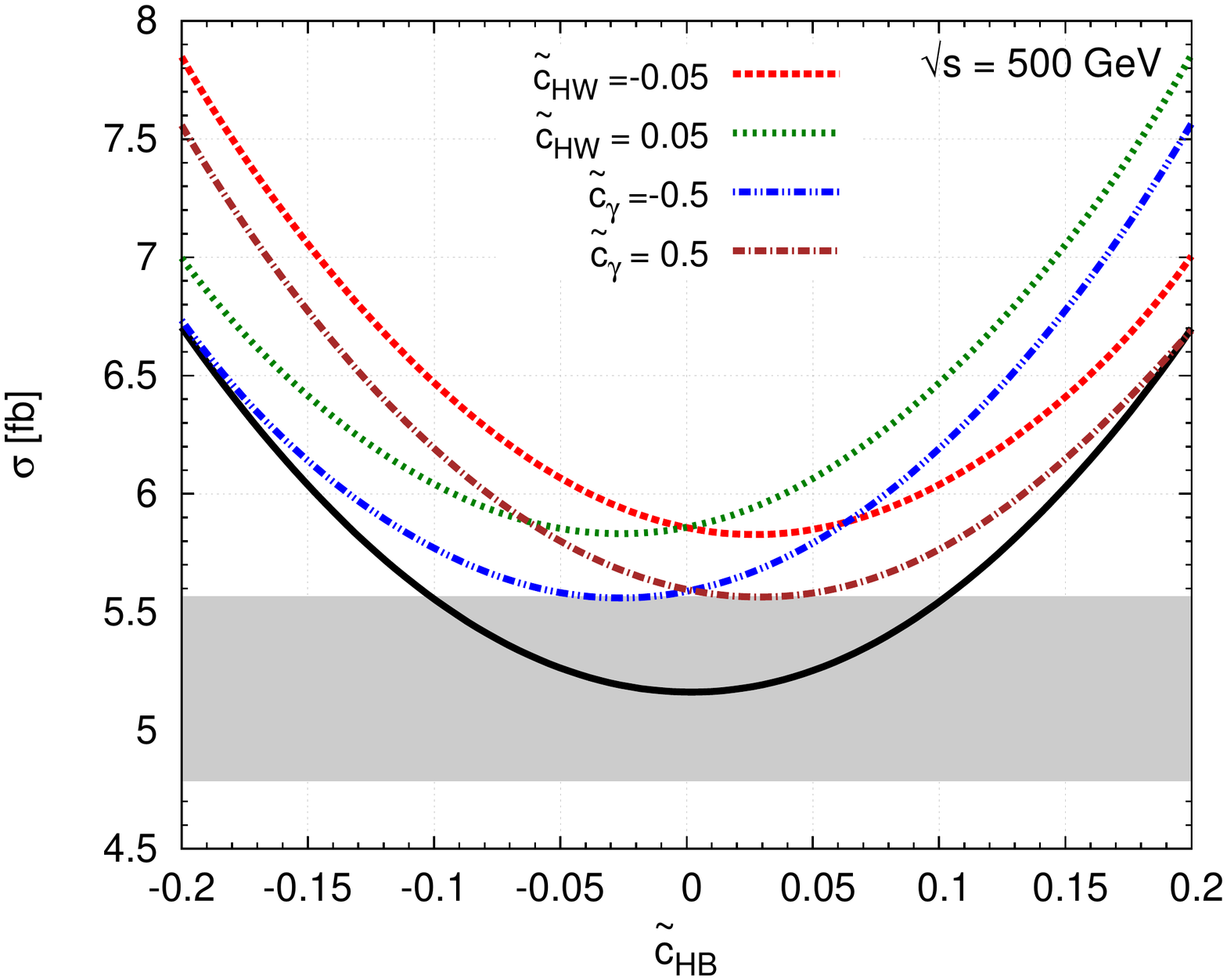} \\
\end{tabular}
\vspace{-10mm}
\caption{Cross section against $\tilde c_{HB}$ in the presence of selected CP-conserving (left) and CP-violating (right) couplings.
The black solid line corresponds to the case when only $\tilde c_{HB}$ is present. The center-of-mass energy is assumed to be 
$\sqrt{s}=500$ GeV. In each case, all other parameters are set to zero. The gray band indicates the $3\sigma$ limit 
of the SM cross section, with an integrated luminosity of $300$~fb$^{-1}$.}
\label{fig:cstcHB}
\end{figure}

\begin{figure}[H] \centering
\begin{tabular}{c c}
\hspace{-10mm}
\includegraphics[angle=0,width=90mm]{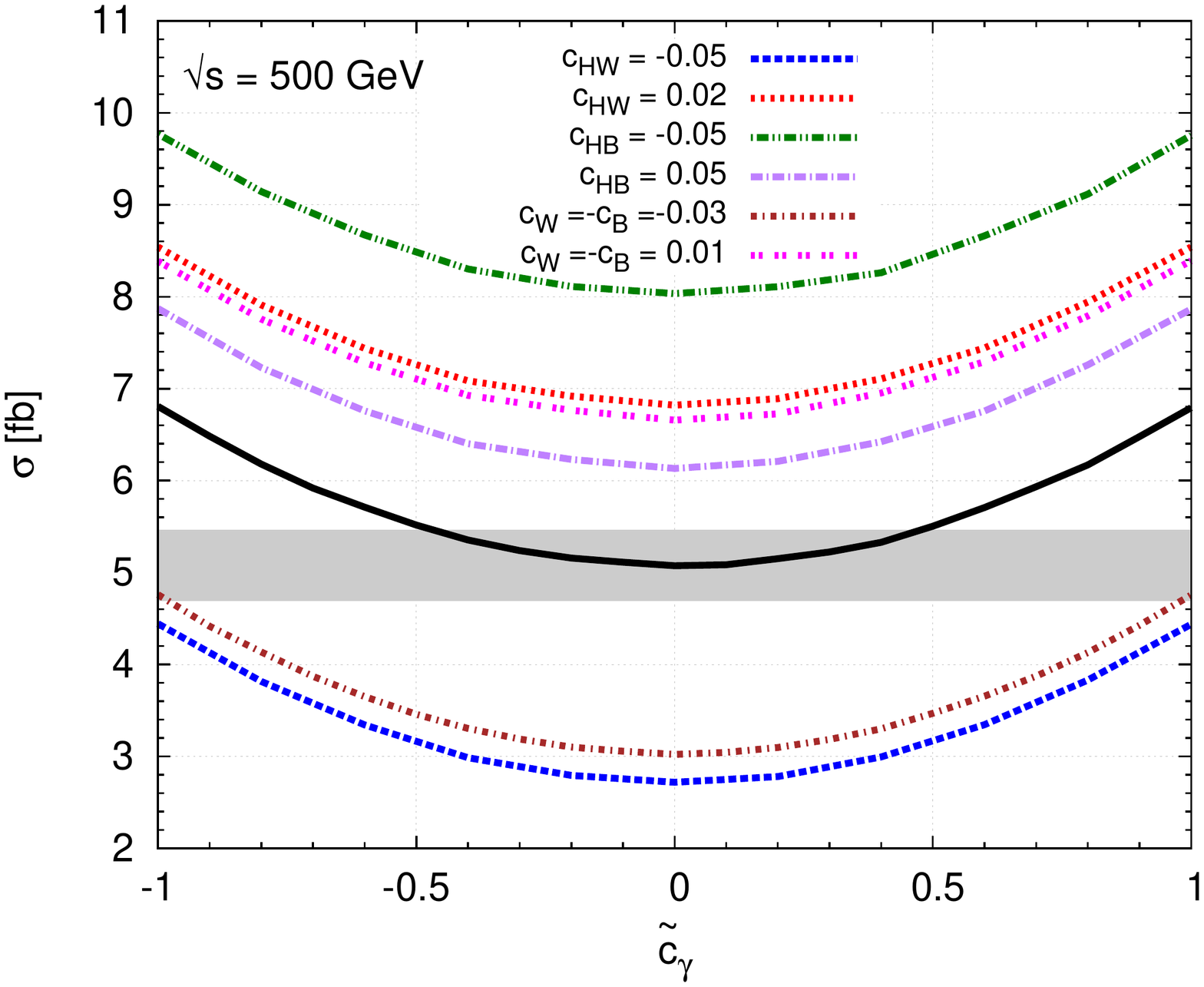} &
\hspace{-16mm}
\includegraphics[angle=0,width=90mm]{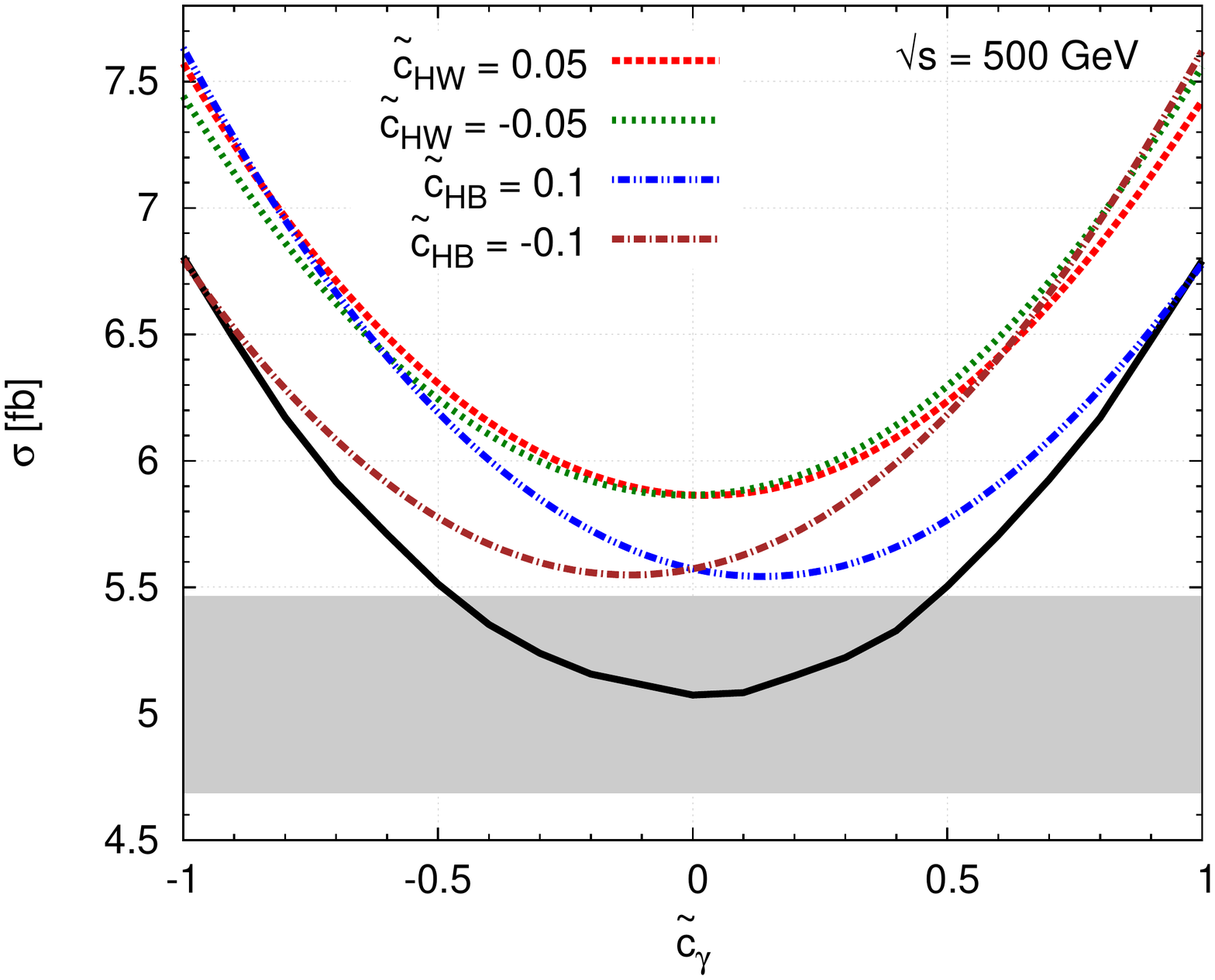} \\
\end{tabular}
\vspace{-10mm}
\caption{Cross section against $\tilde c_{\gamma}$ in the presence of selected CP-conserving (left) and CP-violating (right) couplings.
The black solid line corresponds to the case when only $\tilde c_{\gamma}$ is present. The center-of-mass energy is assumed to be 
$\sqrt{s}=500$ GeV. In each case, all other parameters are set to zero. The gray band indicates the $3\sigma$ limit 
of the SM cross section, with an integrated luminosity of 300~fb$^{-1}$.}
\label{fig:cstcA}
\end{figure}

Coming to the CP-violating couplings $\tilde c_{HW}$, $\tilde c_{HB}$, and $\tilde c_\gamma$, the single parameter reach of the ILC at 500 GeV with $300$~fb$^{-1}$ at the $3 \sigma$ level could be obtained from Figs.~\ref{fig:cstcHW}, \ref{fig:cstcHB}, and \ref{fig:cstcA}, respectively. The effects of other couplings in deriving these limits are also indicated in these figures. Clearly, precise knowledge of the CP-conserving parameters $\bar c_W$, $\bar c_{HW}$ and $\bar c_{HB}$ is required to obtain a reasonably robust estimate of the CP-violating parameters. Among the CP-violating couplings, $\tilde c_{HW}$ affects the cross section most significantly, and the limits derivable on the other parameters are sensitive to their presence. The effect of the $\tilde c_\gamma$ is much smaller than the other couplings in finding the sensitivity of $\tilde c_{HW}$ and, therefore, not presented. 

\begin{figure}[H] \centering
\begin{tabular}{c}
\hspace{24mm}
\includegraphics[angle=0,width=13cm]{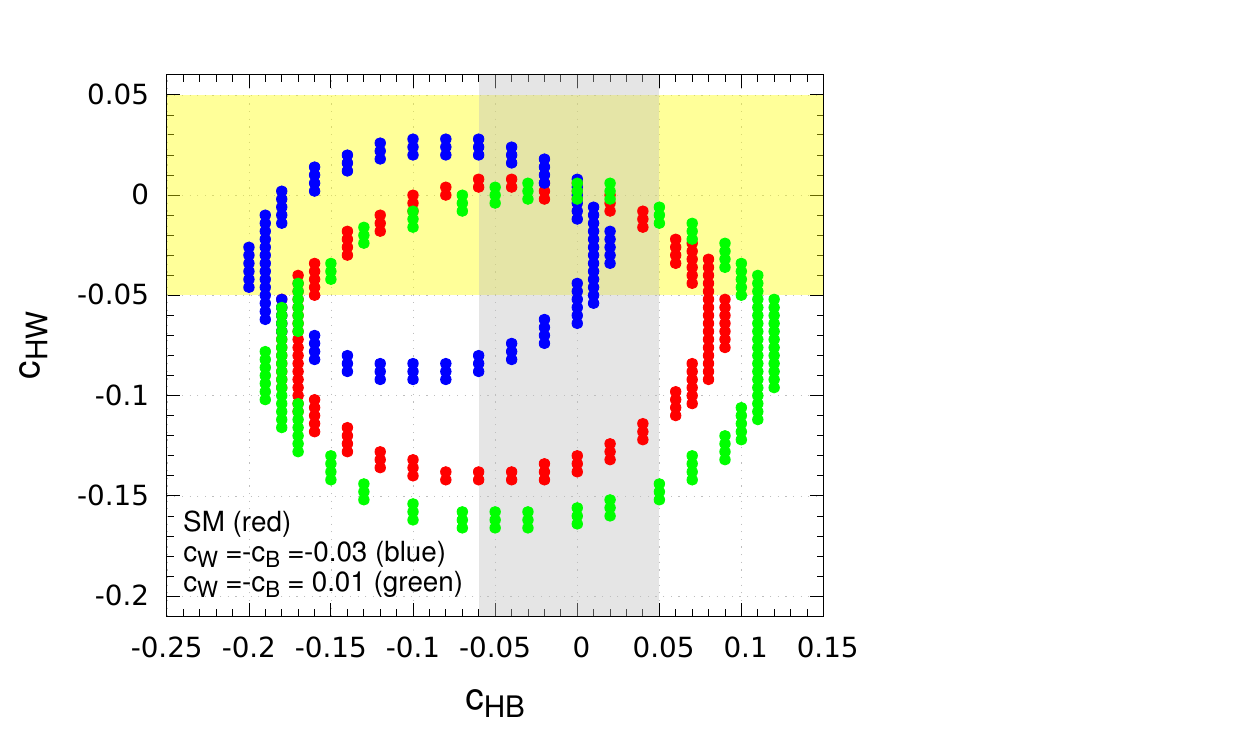}
\end{tabular}
\vspace{-6mm}
\caption{The ellipses correspond to regions in the $\bar c_{HB} -~\bar c_{HW}$ plane with the total cross section within the $3\sigma$ limit of the SM cross section (red), and cross sections with $\bar c_W=-0.03$ (blue) and $\bar c_W=+0.01$ (green). An integrated luminosity of 300~fb$^{-1}$ is considered, and the center-of-mass energy is taken as 500 GeV. The yellow and gray bands correspond to the present limits of $\bar c_{HW}$ and $\bar c_{HB}$, respectively.}
\label{fig:correlation1}
\end{figure}

The correlation between the $\bar c_{HW}$ and $\bar c_{HB}$ is presented in Fig.~\ref{fig:correlation1}, where the yellow and gray bands show the present limits derived from the LHC results on the associated production of the Higgs boson with the $W$ boson \cite{Ellis:2014dva}. In the absence of any other parameter, the allowed region in the $\bar c_{HW} - \bar c_{HB}$ plane is restricted to a narrow ellipse (red). This ellipse is not affected much by the presence of $\bar c_W$ if it is positive (green ellipse). On the other hand, if $\bar c_W$ is negative, within the present bounds, it can significantly affect the allowed region (blue ellipse) in the $\bar c_{HW} - \bar c_{HB}$ plane. The presence of CP-violating parameters is found to be insignificant here.

\begin{figure}[h]\centering
\begin{tabular}{c c}
\hspace{10mm}
\includegraphics[angle=0,width=100mm]{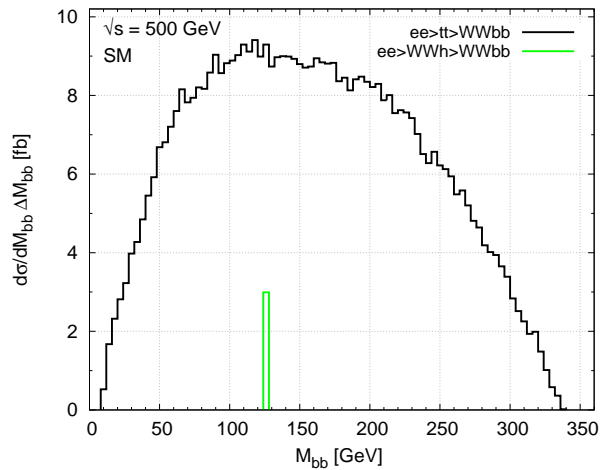} 
\end{tabular}
\vspace{-12mm}
\caption{Invariant mass of $b\bar b$ in the signal (green) and background $t\bar t$ (black) processes at a center of mass energy of $\sqrt{s}=500$ GeV with unpolarized beams.}
\label{fig:background}
\end{figure}

It is essential to know the behavior of various kinematic distributions, and how the anomalous coupling parameters influence these in order to derive any useful and reliable information from the experimental results. This is so, even in cases where the fitting to obtain the reach of the parameters is done with the total number of events, as the reconstruction of events and the reduction of the background depend crucially on the kinematic distributions of the decay products. In the following, we shall present some illustrative cases of distributions at the production level to understand the effect of different couplings on these. 
The dominant decay channel for $h$ in the signal process is $h\rightarrow b\bar b$, with about a 57\% branching fraction. Considering the pure hadronic (with four jets) or semileptonic  (2 jets + lepton + missing energy) decay of the $W$ pair allow one to reconstruct the events. Thus, with the final state as $WWb\bar b$, the $t\bar t$ and $WWZ$ production processes could act as the background. The total cross section of $t\bar t$ and $WWZ$ (with $Z\rightarrow b\bar b$) production processes are about 500 fb and about 6 fb, respectively. Both of these processes could be contained with the help of the invariant mass distribution of the $b\bar b$ pair ($M_{b\bar b}$). In the case of a signal process this is expected to peak at the Higgs mass of about 125 GeV, while in the case of the $WWZ$ process, it is expected to peak around the $Z$ mass of about 91 GeV. Thus, the $WWZ$ background could be taken care of without much trouble. On the other hand, in the case of $t\bar t$ pair production, the $M_{b\bar b}$ distribution is spread out. As presented in Fig. \ref{fig:background}, in the relevant window of $124 - 128$ GeV of the $M_{b\bar b}$, we have about 900 signal events and 2700 background events at 300 fb$^{-1}$ integrated luminosity, leading to a large signal significance of about 15. Of course, this estimate is considering an ideal setting, whereas the reconstruction efficiencies, detector effect, etc. would bring this down considerably. However, one may expect a large significance, even after these realistic considerations.
Presently, we would like to be content with the analysis at the production level, considering the limited scope of this work. As mentioned earlier, we shall focus on an ILC running at a center-of-mass energy of 500 GeV for our study. In order to understand  the interplay of CP-conserving and CP-violating couplings, we consider $\bar c_W$ and $\bar c_{HW}$ only with the CP-violating couplings. 

\begin{figure}[H] \centering
\begin{tabular}{c c}
\hspace{-10mm}
\includegraphics[angle=0,width=90mm]{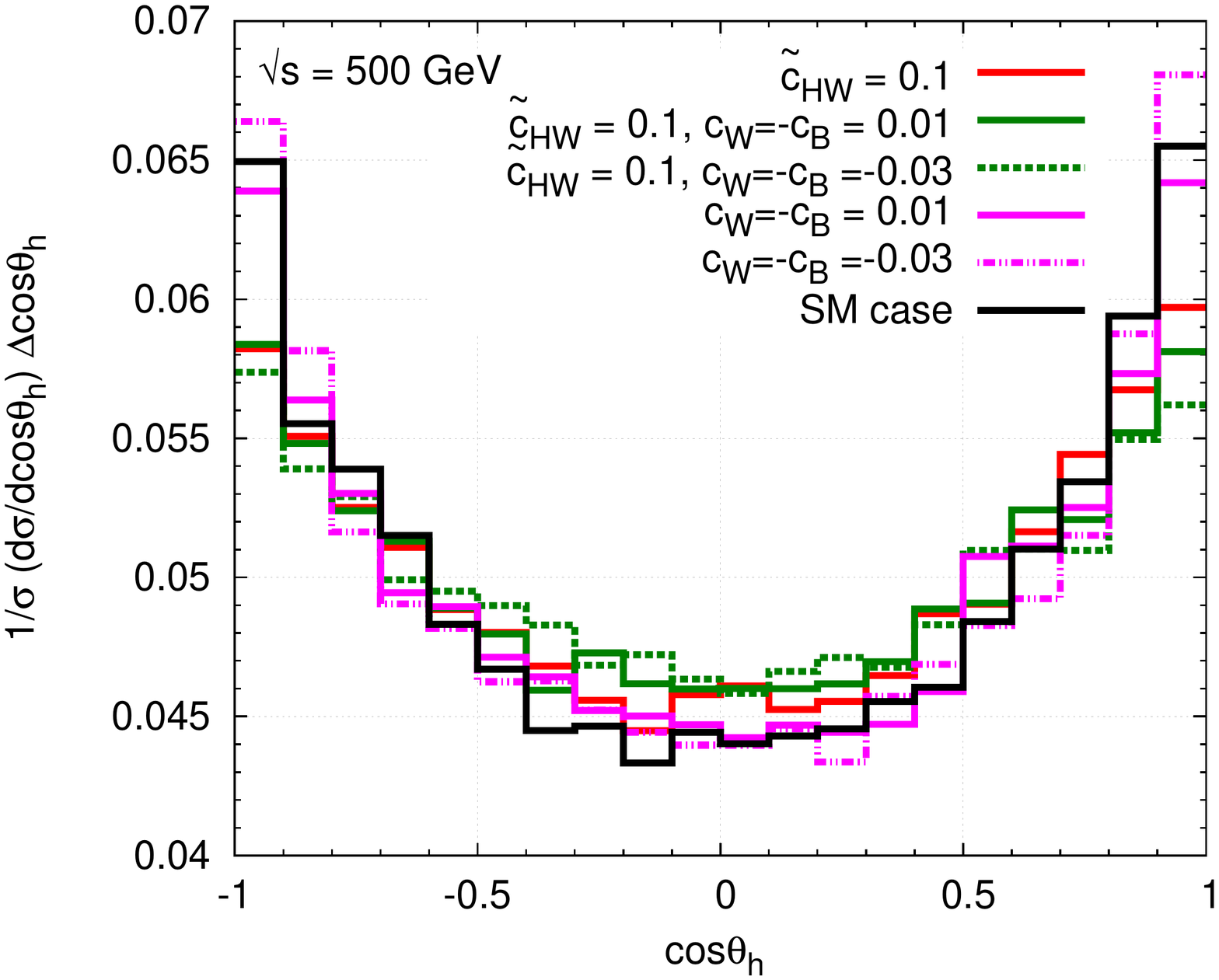} &
\hspace{-16mm}
\includegraphics[angle=0,width=90mm]{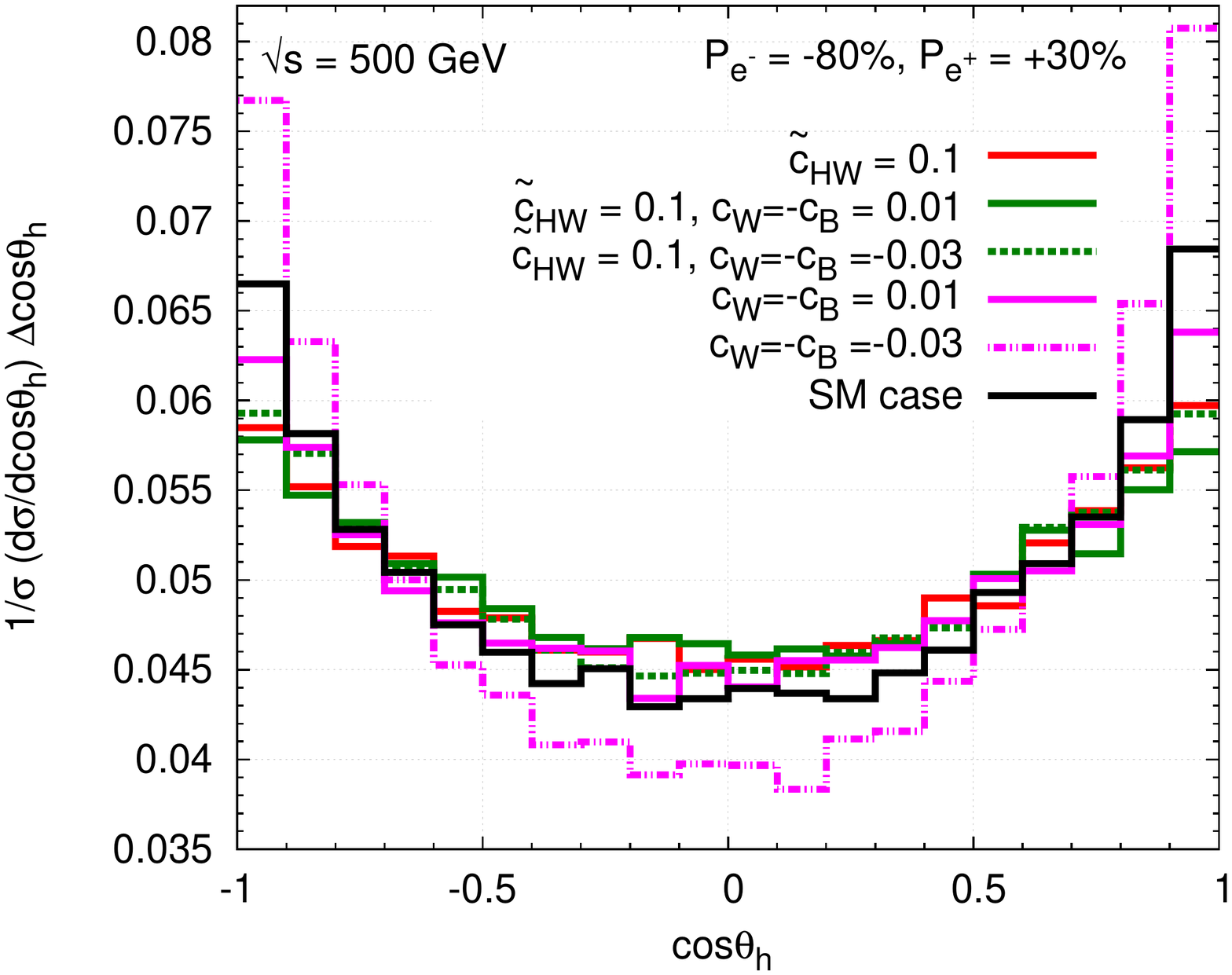} \\
\end{tabular}
\vspace{-12mm}
\caption{Distribution of $\cos\theta_h$ for different anomalous couplings with unpolarized (left) and polarized with $P_{e^-}=-80\%,~~P_{e^+}=+30\%$ (right) beams. A center-of-mass energy of $500$ GeV is assumed.}
\label{fig:angH}
\end{figure}

\begin{figure}[H] \centering
\begin{tabular}{c c}
\hspace{-10mm}
\includegraphics[angle=0,width=90mm]{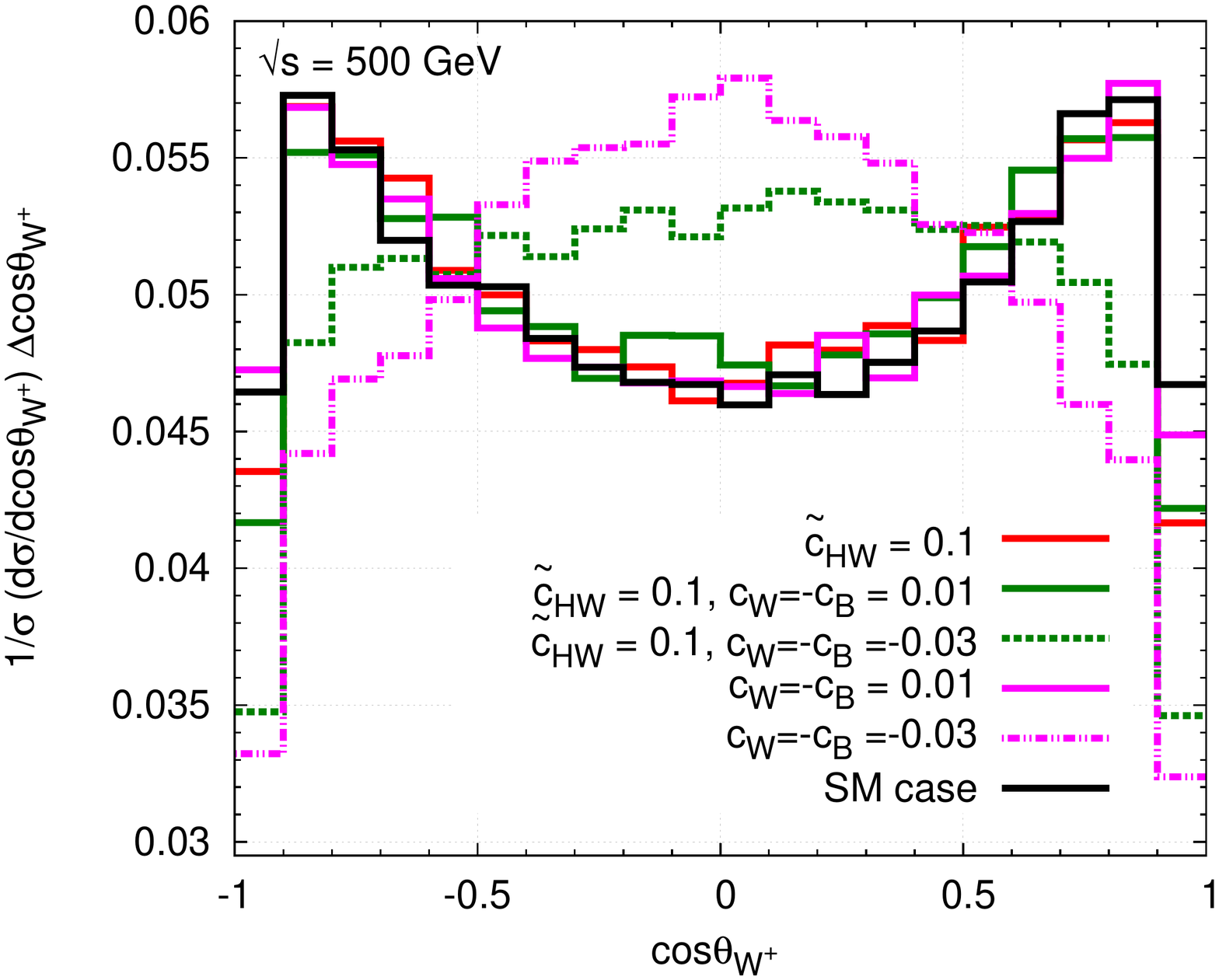} &
\hspace{-16mm}
\includegraphics[angle=0,width=90mm]{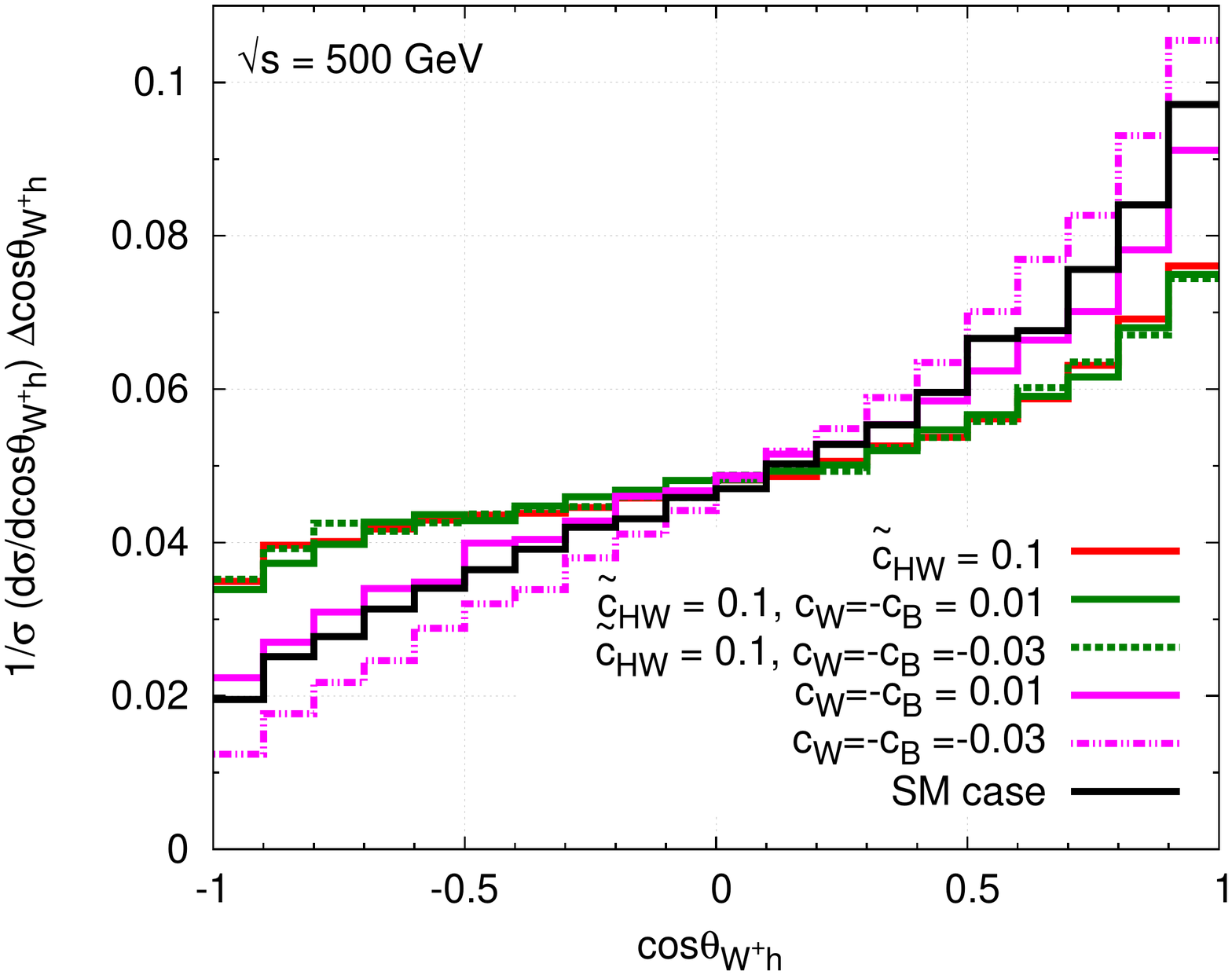} \\
\end{tabular}
\vspace{-12mm}
\caption{Distribution of $\cos\theta_{W^+}$ (left) and $\cos\theta_{W^+h}$ (right) for different anomalous couplings with unpolarized beams. A center-of-mass energy of $500$ GeV is assumed.}
\label{fig:angW}
\end{figure}

\begin{figure}[H] \centering
\begin{tabular}{c c}
\hspace{-8mm}
\includegraphics[angle=0,width=90mm]{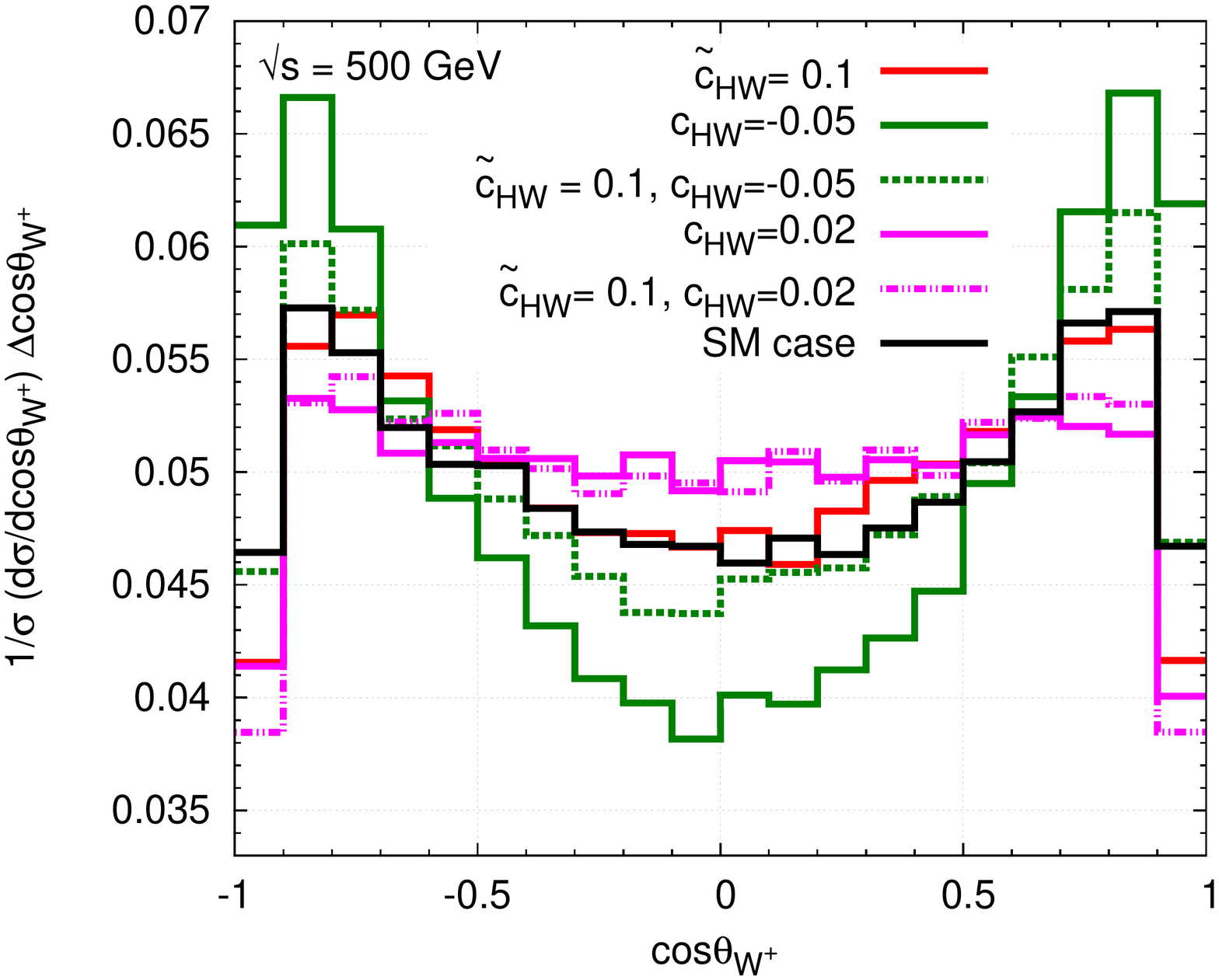} &
\hspace{-16mm}
\vspace{-8mm}
\includegraphics[angle=0,width=90mm]{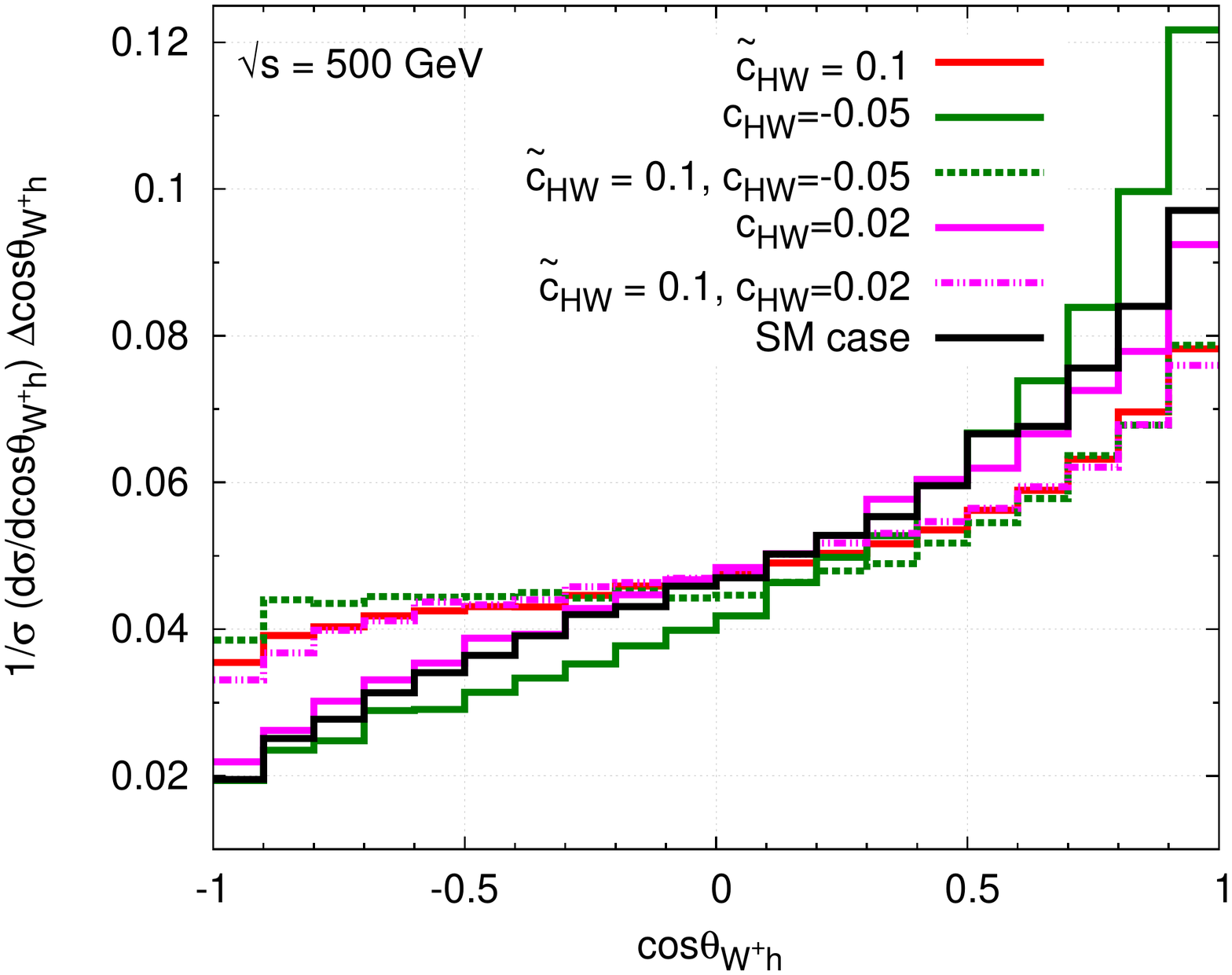} 
\end{tabular}
\vspace{-3mm}
\caption{Distribution of $\cos\theta_{W^+}$ (left) and  $\cos\theta_{W^+h}$ (right) for different anomalous coupling values. A center-of-mass energy of $500$ GeV is assumed. The color coding is the same in both figures.}
\label{fig:angHW_cHW}
\end{figure}

\begin{figure}\centering
\begin{tabular}{c c}
\hspace{-8mm}
\includegraphics[angle=0,width=90mm]{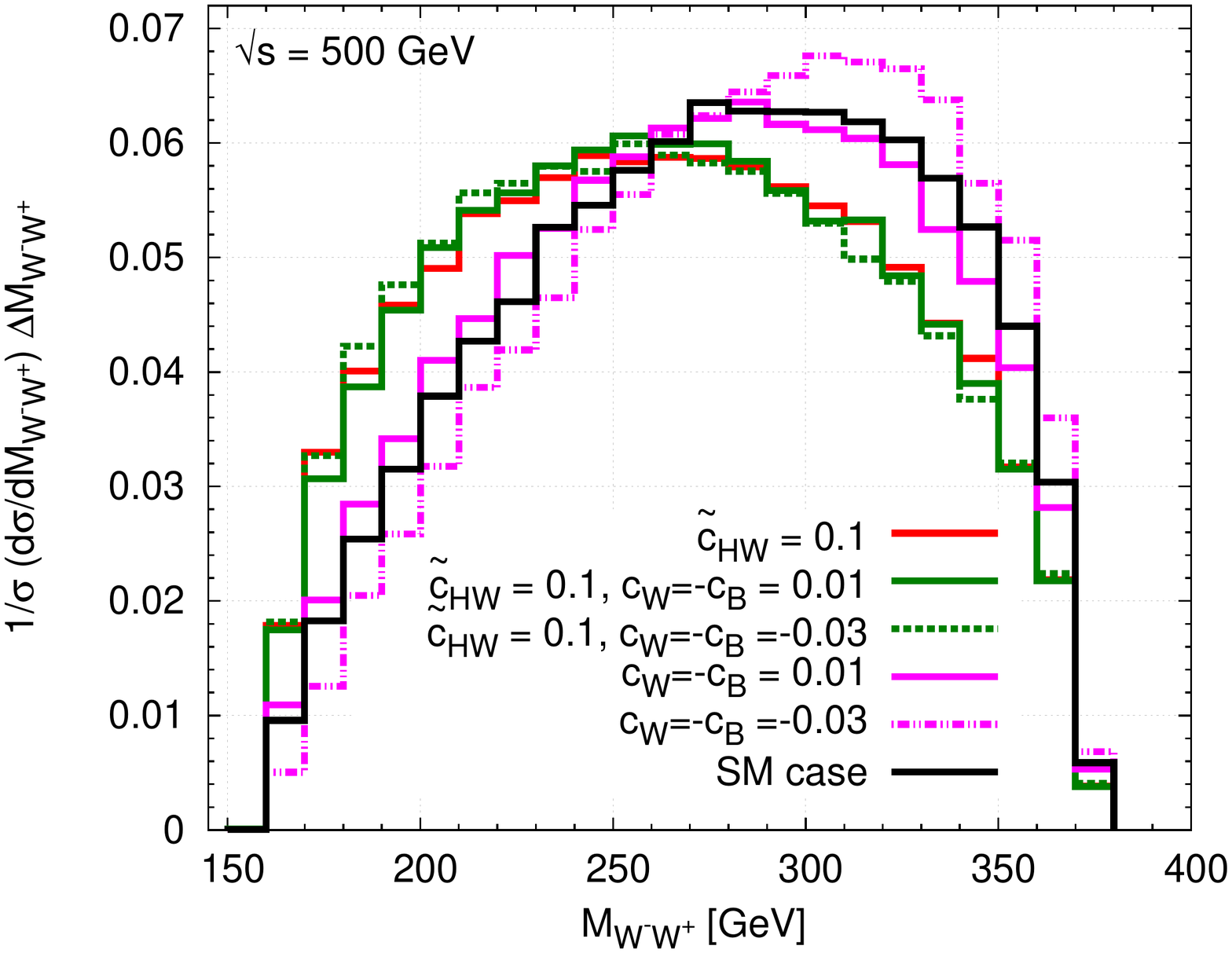} &
\hspace{-16mm}
\vspace{-9mm}
\includegraphics[angle=0,width=90mm]{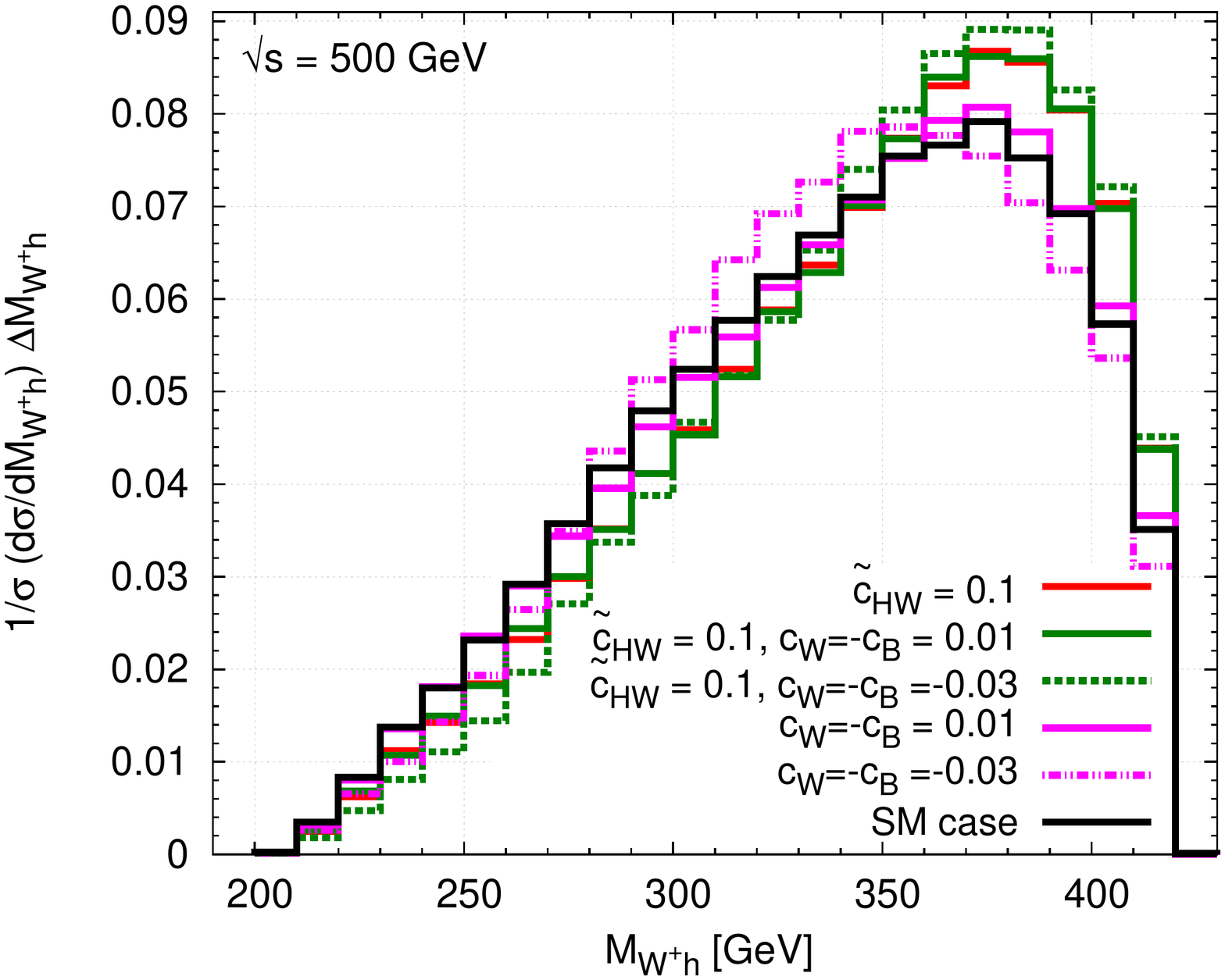}
\end{tabular}
\vspace{-3mm}
\caption{The invariant mass distribution of $W^-W^+$ (left)  and  $W^+h$ (right) for different anomalous coupling values. A center-of-mass energy of $500$ GeV is assumed. The color coding is the same as in Fig. \ref{fig:angH}.}
\label{fig:invmass_HW}
\end{figure}

The effect of the anomalous couplings on the kinematic distributions are presented in Figs. \ref{fig:angH} - \ref{fig:invmass_HW}.  The couplings $\tilde c_{HB}$ and $\tilde c_\gamma$ are found to have no significant effect, with or without the presence of other parameters and, therefore, are not presented here. In the figures, the couplings other than those mentioned are set to zero. The parameters having a significant effect are the $CP$-violating couplings $\tilde c_{HW}$ and the $CP$-conserving couplings $\bar c_W$ and $\bar c_{HW}$. The parameter choices considered for these numerical analyses are \\
\begin{center}
\(
\bar c_W= -0.03, +0.01, ~~~~~\bar c_{HW}=-0.05, +0.02, ~~~~~\tilde c_{HW}=0.1
\)
\end{center}

\noindent
While for $\bar c_W$, the maximum allowed values as per the present bounds are used, in the case $\bar c_{HW}$, it is somewhat arbitrary but within the limits. In the case of the $CP$-violating parameter $\tilde c_{HW}$, no such limits exist, and we have considered a conservative choice of an arbitrary value to illustrate its influence. Unlike the case of $\bar c_W$ [as seen in Fig. \ref{fig:cs1} (right)], the sign of $\tilde c_{HW}$ is irrelevant, as seen in the symmetric plots in Fig. \ref{fig:cstcHW}. While considering beam polarization, an 80\% left-polarized electron beam and a 30\% right-polarized positron beam are assumed, as is expected in the first phase of the ILC, according to the present baseline design. 

We first consider in Fig.\ref{fig:angH} the normalized $\cos\theta_{h}$ distributions of the Higgs boson for the SM case, as well as different cases with anomalous couplings (both CP conserving and violating) as indicated in the figure, while all other parameters are set to zero.  The normalized distributions provide clear information on the shape of the distribution, bringing out the qualitative difference between the different cases considered.  The shape of the distribution remains more or less the same as that of the SM case, except a small enhancement in the central regions when both the couplings are nonzero (green curves). The advantage of beam polarization is evident (figure on the right) when compared to the corresponding unpolarized (figure on the left) case. Here, the case of negative $\bar c_W$ differs from the other cases. This feature can be exploited to discriminate this case from others.

Figure \ref{fig:angW} (left) presents the normalized $\cos\theta_{W^+}$ distribution (unpolarized beams). The negative value of $\bar c_W$ changes the nature of the distribution drastically (dashed magenta) compared to the SM case (solid black), while all other cases have insignificant deviation. This again can be a useful discriminator of the case, but unlike the case of $\cos\theta_h$ distribution, visible deviation is present even in the presence of nonzero $\tilde c_{HW}$. The presence of beam polarization leaves the shape of the distribution largely unchanged. At the same time, as seen earlier, the cross section itself is enhanced by a factor of a little more than 2.
Figure \ref{fig:angW} (right) shows the normalized $\cos\theta_{W^+h}$ distribution (unpolarized beams), where $\theta_{W^+h}$ is the angle between $h$ and $W^+$. Here, $\tilde c_{HW}$ has significant effect, which is not affected by the presence of $\bar c_W$. Thus, an enhancement in the backward region and a corresponding decrease in the forward region compared to the SM case indicate nonzero $\tilde c_{HW}$. 
On the other hand, the presence of negative $\bar c_W$ (dashed magenta) alone has the opposite effect. 
This along with $\cos\theta_{W^+h}$ will be able to fix the case between the presence of $\tilde c_{HW}$, $\bar c_W$ alone or together. Here again, it is seen that the use of polarized beams, while helping improve the statistics, keeps the qualitative features unchanged. Considering these three angular distributions together might allow us to distinguish different scenarios. For example, if $\tilde c_{HW}$ alone is present, we may expect a significant effect in the $\cos\theta_h$ and the $\cos\theta_{W^+h}$ distributions, whereas $\cos\theta_{W^+}$ distribution remains more or less unaffected. Along with $\tilde c_{HW}$, if $\bar c_W$ was present (either positive or negative), the effect in $\cos\theta_h$ is nullified, whereas the effect would remain in $\cos\theta_{W^+h}$. The change in $\cos\theta_{W^+}$ as shown in Fig.\ref{fig:angW} (left) indicates the presence of the negative value of $\bar c_W$ with or without the presence of other couplings. Table~\ref{table:angle-distinguish} summarizes the cases that could be distinguished.

\begin{table}[H]
\begin{center}
\begin{tabular}{|l|c|c|c|}
\hline
Couplings&$\cos\theta_h$&$\cos\theta_{W^+}$&$\cos\theta_{W^+h}$\\ \cline{1-4}
&&&\\
$\tilde c_{HW}$ alone&Yes&No&Yes\\
$\bar c_W$ (positive) alone &No&No&No\\
$\bar c_W$ (negative) alone&Yes&Yes&Yes\\
$\tilde c_{HW}$ and $\bar c_W$ (positive)&No&No&Yes\\
$\tilde c_{HW}$ and $\bar c_W$ (negative)&No&Yes&Yes\\[5mm]
\hline
\end{tabular}
\caption{Presence (Yes) or absence (No) of deviations that could be expected in case of different scenarios with combinations of $\bar c_W$ and $\tilde c_{HW}$ realized from Figs.~\ref{fig:angH} and \ref{fig:angW}.}
\label{table:angle-distinguish}
\end{center}
\end{table}

Figure \ref{fig:angW} suggests that the forward-backward asymmetry is a quantitative estimator of the presence of anomalous couplings. The percentage of deviation from the SM case for the cases of a considered set of parameters at fixed center-of-mass energy of $500$ GeV without and with polarized beams is given in Table \ref{table:asymmetry500}, where the asymmetry is defined as 

\begin{equation}
{A_{FB} = \frac{\left[\int_{-1}^{0}\frac{d\sigma}{d\cos\theta}d\cos\theta - \int_{0}^{1}\frac{d\sigma}{d\cos\theta}d\cos\theta\right]}{\left[\int_{-1}^{0}\frac{d\sigma}{d\cos\theta}d\cos\theta + \int_{0}^{1}\frac{d\sigma}{d\cos\theta}d\cos\theta\right]}}
\label{eqn:FBc_asymmetry}
\end{equation}
\begin{equation}
\Delta A_{FB}(\%) = \frac{\left|A_{FB}^{ano} - A_{FB}^{SM}\right|}{A_{FB}^{SM}}\times100.
\label{eqn:percentageasymmetry}
\end{equation}
\begin{table}[H]
\begin{center}
\begin{tabular}{|c|c|c|c|}
\hline
$\tilde c_{HW}$&$\bar c_W=-\bar c_B$ &\multicolumn{2}{c|}{\textbf{$\Delta A_{FB}(\cos\theta_{W^+h}) \%$}}  \\\cline{3-4} 
&&Unpolarized beams &$P_{e^-}=-80\%$,  $P_{e^+}=30\%$   \\ 
\hline\hline
0.1   &$~~0$      &$50$   &$53$     \\\cline{1-4}
0.1   &$~~0.01$   &$52$   &$52$     \\\cline{1-4}
\hline
0.1  &$-0.03$  &$52$   &$63$     \\\cline{1-4}
\hline
0     &$~~0.01$   &$13$   &$11$     \\\cline{1-4}
\hline
0    &$-0.03$  &$31$   &$43$     \\\cline{1-4}
\hline
\hline
\multicolumn{2}{|c|}{SM~case;  $A_{FB}=$}    &$0.3117$    &$0.3102$    \\\cline{1-4}
\hline
\end{tabular}
\caption{ {Observed forward-backward asymmetry and its deviation from the SM in the  angular distribution ($\theta_{W^+h}$) at center-of-mass energy of $500$ GeV.}}
\label{table:asymmetry500}
\end{center}
\end{table}

The case of $\bar c_{HW}$ along with CP-violating parameters also presents a similar picture. In Fig.~\ref{fig:angHW_cHW}, we present $\cos\theta_{W^+}$ and $\cos\theta_{W^+h}$ as an example ($\bar c_{HB}$ is found to be less sensitive).  Here again the influence of $\bar c_{HW}$ on the sensitivity of $\tilde c_{HW}$ is clear. Similar features are also present in other kinematic distributions like $\cos\theta_h$ and $\cos\theta_{W^+}$. Unlike the case of $\bar c_W$ (presented in Figs. \ref{fig:angH} and \ref{fig:angW}), here we do not find possibilities to distinguish different scenarios with the help of these distributions.

Finally, we consider the normalized invariant mass distributions of $W^-W^+$ and $W^+h$. Figure \ref{fig:invmass_HW} presents the sensitivity of invariant mass distribution to the anomalous coupling parameters for the same set of parameters as in the inset of Fig.\ref{fig:angH}. The combinations of the parameters affected are similar to the case of $\cos\theta_{W^+h}$. This can thus, provide an additional tool to distinguish these scenarios. 

Note that in all cases, the beam polarization is found to be useful in terms of improved sensitivity with more than double the number of events compared to the case of the unpolarized beam, while keeping the qualitative features (shape of the curve) unaffected (except in the case of $\cos\theta_h$ distributions, where the shape is different in the case of $\bar c_{W}=-0.03$). Thus, the reach of the probe of the couplings can be improved by a factor of 1.5 to 2 in all cases.  
 
\section{Summary and Conclusions} \label{sec:summary}
The discovery of the Higgs boson by the ATLAS and CMS collaborations at the LHC has confirmed the Higgs mechanism as the way to have EWSB providing masses to the fundamental particles. The properties of the Higgs boson measured by the LHC so far are consistent with the expectations of the SM. It is expected that the LHC would measure the mass, spin, and parity of this particle along with the standard decay widths somewhat precisely. On the other hand, details of the couplings like the trilinear and quartic self-couplings as well as the couplings with the gauge bosons are not expected to be measured precisely. At the same time, precise knowledge of these couplings is very important in reconstructing the EWSB mechanism. A precision machine like the ILC is expected to help in the precise measurement of these couplings. In this paper, the process $e^-e^+ \rightarrow W^-W^+h$, which is influenced by the Higgs to gauge boson couplings, namely, $WWh,~WW\gamma$, and $ZZh$, is considered. The reach of an ILC at $\sqrt{s}=500$ GeV with an integrated luminosity of 300~fb$^{-1}$ in probing the different relevant parameters of the corresponding effective Lagrangian is presented. The influence of the presence of other couplings in the probe of each of the couplings is studied. In general, it is observed that the CP-violating coupling $\tilde c_\gamma$ has very small effect on almost all of the observables considered. Study of the $\bar c_{HW}-\bar c_{HB}$ plane shows that the allowed region can be narrowed to a very small band. While this band is unaffected by the presence of $\bar c_W>0$, the effect is significant if $\bar c_W<0$.  Consideration of the angular distributions of the Higgs boson ($\cos\theta_h$), the $W$ boson ($\cos\theta_{W^+}$) and the distributions of the angle between $W^+$ and $h$, ($\cos\theta_{W^+h}$) proves to provide a handle in distinguishing the presence of different combinations of $\bar c_W$ and $\tilde c_{HW}$. All other parameters have an indistinguishable effect on these distributions. The invariant mass distributions of the $W^-W^+$ pair as well as the $W^+h$ pair are also sensitive to some combinations of the above parameters. A quantitative estimate of the forward-backward asymmetry corresponding to the angle between $W^+$ and $h$ shows that large deviations of up to 50\% are possible for moderate values of the couplings. In all cases, a suitably chosen beam polarization is found to be advantageous, as illustrated with an 80\% left-polarized electron beam and a 30\% right-polarized positron beam. 
The statistics can be improved by a factor of 2 with the baseline polarization quoted above, which can be improved to an enhancement factor of 3 with the expected 60\% positron beam polarization in the upgraded version of the ILC.  This, along with the fact that the qualitative features of the distributions in almost all cases are kept more or less intact, can be used to improve the reach of the ILC in probing these couplings significantly (almost a factor of 2). In at least one case of $\cos\theta_h$ distribution, we notice a qualitative change in the case of nonzero $\bar c_W$ when all other couplings are absent. Apart from the overall normalizing factor, some details are also affected, as is illustrated in the improvements in the forward-backward asymmetry, when beam polarization is used. Thus, 
the study shows that $WWh$ production at the ILC is useful in detecting the anomalous couplings in Higgs-gauge boson interactions. A detailed analysis involving standard kinematic distributions could be used to distinguish different scenarios involving the couplings. While the numerical study needs to be improved with more realistic collider and detector information, as well as study of the background processes, we hope to have conveyed the importance of the process in determining and disentangling the effects of anomalous Higgs-gauge boson couplings.

\vskip 5mm
\noindent
{\bf Acknowledgments} \\[2mm]
We would like to thank Dr. Sumit Garg for useful discussions and technical help.

\end{document}